\documentclass[preprint,12pt]{elsarticle}
\usepackage{amssymb,amsfonts,amsthm}
\usepackage{geometry}
\usepackage{graphics,subfigure}
\usepackage{hyperref}
\usepackage{pifont}
\usepackage{natbib}
\biboptions{sort&compress}
\usepackage[fleqn]{amsmath}
\usepackage{color}
\usepackage{lineno}
\usepackage{bm}

\begin{document}

\title{An experimental study of morphological formation in bilayered tubular structures driven by swelling/growth}

\author[add1]{Rui-Cheng Liu}
\author[add2]{Lishuai Jin}
\author[add1,add3]{Zongxi Cai}
\author[add1,add3]{Yang Liu\corref{cor1}}
\ead{tracy\underline{ }liu@tju.edu.cn}

\address[add1]{Department of Mechanics, School of Mechanical Engineering, Tianjin University, Tianjin 300354, China}
\address[add2]{Department of Materials Science and Engineering, University of Pennsylvania, Philadelphia, PA 19104, USA}
\address[add3]{Tianjin Key Laboratory of Modern Engineering Mechanics, Tianjin 300354, China}

\cortext[cor1]{corresponding author}

\begin{abstract}
Circumferential wrinkling in soft tubular tissues is vital in supporting normal physiological functions. Most existing literature was dedicated to theoretical modelling and finite element simulations based on a specific growth model. This paper presents an experimental investigation on pattern formation and evolution in bilayered tubular organs using swelling deformation of polydimethylsiloxane (PDMS) and aims at supplying a thorough comparison with theoretical and finite element results. To create a twin model in modelling and simulation, the shear modulus in the incompressible neo-Hookean material is estimated via uni-axial tensile and pure shear tests. Five bilayered tubes with different material or geometrical parameters are fabricated. Swelling experiments are carried out for these samples in an individual experimental setup where a plane-strain deformation is guaranteed, and several surface patterns and the associated mode transformations are observed, namely, creases, wrinkles, period-doubling profiles, wrinkle-to-crease transition, and wrinkle-to-period-doubling transition. In particular, an interfacial wrinkling pattern is also observed. To make comparisons, a buckling analysis is conducted within the framework of finite elasticity by means of the Stroh formulation and a refined surface impedance matrix method. In addition, a finite element analysis is performed to trace the evolution of surface instabilities. It turns out that the experimental findings agree well with the theoretical predictions as well as the finite element results. From our experiments, it is found that creasing mode may appear instead of wrinkling mode when both layers share a similar mechanical property. It is expected that the current work could provide novel experimental insight into pattern formation in tubular structures. In particular, the traditional impedance matrix method has been adapted, which enables us to resolve eigenvalue problems with displacement boundary conditions, and the good agreement among experimental, theoretical, and simulation consequences supplies strong evidence that a phenomenological growth model is satisfactory to reveal mechanisms behind intricate surface morphology in tubular tissues. 
\end{abstract}
\maketitle

\section{Introduction}
Multiple surface patterns in soft materials and biological tissues induced by external stimuli or spontaneous growth, such as wrinkles, creases, and folds, have unfolded enormous applications in flexible electronics \cite{1,2,3,4}, surface engineering \cite{5,6}, morphogenesis of plants and animal tissues \cite{7,8,9,10}, and biomedical engineering \cite{11,12,13}. In general, a specific surface pattern in soft tissues will play a pivotal role in the development of organisms and in determining the standard physiological functions. Any abnormal growth or remodeling may affect normal morphology and then generate pathological diseases in turn. For instance, mucosal growth associated with thickening of smooth muscle surrounding an airway will trigger mechanical instabilities that result in inward folding and airway obstruction, which could cause chronic lung disease \cite{14}, and brain gyri and sulci are closely relevant to neurological dysfunction and a thickened cortex will suppress folding formation and further produce lissencephaly \cite{15}. It must be pointed out that the surface of a soft tissue is smooth in its earlier stage and pattern formation may be triggered at a critical occasion due to constrained  growth. This prominent feature is analogue to classical Euler buckling instability \cite{16}. Also, growth-induced patterns can be viewed as a consequence of solution bifurcation in the growing process. It is therefore of fundamental significance to elucidate the role of various geometrical and physical parameters as well as growth process in regulating the final patterns. To this end, Rodriguez et al. proposed a methodology that decomposes the deformation gradient by the multiplier of an elastic deformation gradient and a growth tensor \cite{17}. Later, Goriely and Ben Amar \cite{18} established a growth theory, which is suitable for finite deformation, for soft tissues, and offered some elemental case studies. Since then, much effort was dedicated to supplying useful insight into pattern formation and evolution within the framework of nonlinear elasticity, and we refer to Li et al. \cite{19} and a monograph by Goriely \cite{20} for an exhaustive review.

Of particular interest in this paper is the extensively observed wrinkled morphology in tubular organs, such as artery \cite{21}, bronchus \cite{22}, and gastrointestinal tract \cite{23}. In principle, tubular tissues are generally composed of muscular, submucosal and mucosal layers \cite{24}. Specifically, the muscular layer is usually much stiffer than the other two layers  and can be treated as a rigid confinement in a deformation induced by volumetric growth \cite{23}. Note that volume and mass increase is mainly responsible for soft tissue development and further drives the evolution of surface patterns. Although a practical growth process is extremely complicated combining both biological and chemical factors, a pure mechanical model can still capture the central characteristics that dominate pattern formation and reveal the mechanisms behind various modes using analytical or numerical methods \cite{24,25,26,27,28,29,30,31,32}. Focusing on pattern formation in the circumferential direction, Li et al. \cite{24,25} performed a thorough analysis on the initiation and evolution of growth-induced surface wrinkling in tubular tissues  where the outer surface is fixed using both the theoretical and finite element approaches. Meanwhile, Moulton and Goriely \cite{26,27} studied surface wrinkling in growing cylindrical tubes subject to multiple boundary conditions with applications to asthma. In addition, Ciarletta and Ben Amar \cite{28} proposed a variational approach to identify the wrinkling threshold, and they further unraveled the influence of material anisotropy on both hoop wrinkling and longitudinal wrinkling \cite{29}. Subsequently, Balbi and Ciarletta \cite{30} extended the bifurcation analysis to the bi-directional case. By means of finite element analysis, pattern selection and evolution in bilayered tubular tissues with differential growth were unraveled by Ciarletta et al. \cite{31} and Balbi et al. \cite{32}, respectively. We point out that a linear bifurcation analysis will lead to an eigenvalue problem of an ordinary differential equation with variable coefficients, which makes the pursuing of an explicit bifurcation condition particularly challenging. Jin et al. \cite{33,34} deduced an asymptotic solution for the critical growth ratio and the associated wavenumber based on the Wentzel-Kramers-Brillouin (WKB) technique and presented a semi-analytical framework for deriving the amplitude equation of a single wrinkling mode. Recently, the effect of material inhomogeneity on buckling pattern was investigated and it was found that different modulus gradient can modulate pattern transition \cite{35}.

The above-mentioned studies were mainly concerned with theoretical modelling or numerical simulations and have offered useful insight into pattern formation in soft tubular tissues, especially the influence of geometrical and physical properties on regulating the eventual morphology. In fact, control of biological growth is almost unrealistic, so we only expect to interpret underlying mechanisms for the emergence of abnormal patterns. Then an inherent question arises, how to qualitatively and quantitatively evaluate the robustness and accuracy of a theoretical or numerical prediction?  A proper physical experiment can serve as a benchmark. Although an \textit{in vivo} experiment on biological growth is extremely complicated, swelling deformation of hydrogel or polydimethylsiloxane (PDMS) indeed furnishes an appropriate  paradigm of tissue growth \cite{36,37}. In particular, these polymer  materials have the advantages of low cost, good permeability, and can have large volume changes \cite{38}. On the one hand, Dervaux et al. \cite{39} employed a hydrogel coated to a non-swelling gel immersed into distilled water to model tumor growth. On the other hand, polymer gels such as silicone rubber or PDMS were used to cast brain phantom for the purpose of exploring complex morphology \cite{8,40,41}. For tubular tissues suffering constrained growth, Du et al. \cite{42} designed an experiment of swelling hydrogel to illuminate the role of initial residual stress in tuning pattern selection. Yet a systematic study concerning comparisons between experimentally observed pattern and theoretical or numerical prognostication is still lacking for growing tubular tissues, and this motivates the current work. As a result, we refer to the experimental setup in \cite{8} and harness the swelling property of PDMS when placed in a container filled with hexanes. In particular, the geometrical and physical parameters used in theoretical and numerical models are identical to those of the fabricated samples. To this end, an exhaustive parameter characterization test will be conducted as well.
 
In practice, to produce a physical model of growing bilayered tubular tissues where the outer surface is fixed, we utilize a fabricated simulacrum made of a curved PDMS layer coated to another curved PDMS substrate. Then we inject hexanes into the hollow part to generate swelling deformation. In this way, the inner surface will absorb hexanes first and initially the position far from the inner surface is free of dilation. This setup actually results in a growth gradient decaying from the inner surface. We point out that in our previous analysis the influence of growth gradient on surface wrinkling and pattern transition in growing tubular tissues were unveiled, and it was found that a homogeneous growth field can certainly capture the main features of deformation and instability \cite{43}. This implies that we can ignore the growth gradient, which will facilitate the theoretical analysis and finite element simulations in the present work. In addition, this work can be regarded as a continuation of our previous studies for pattern formation of growing tubular organs, as shown in \cite{33,34,35}, where the bifurcation condition stemming from the eigenvalue problem with variable coefficients was numerically solved using the determinant method or the compound matrix method \cite{44,45}. However, for layered structures where the traction and displacement are continuous across each interface, the Stroh formulation \cite{46} and impedance method \cite{47,48,49}, which have been widely applied in bifurcation analysis of Euler-type buckling \cite{16}, growth-induced pattern formation \cite{10,22}, torsion instability in soft cylinders \cite{50}, and instability of dielectric elastomers \cite{51,52,53}, will enable buckling analysis to be carried out in a compact and subtle fashion. So we adopt this methodology to re-visit the bifurcation analysis. Furthermore, a finite element model of growth is built in Abaqus to invesigate the post-buckling behaviors.

The paper is structured as follows. In Section 2, we introduce the swelling experiments and the material characterization test, and the associated buckling patterns are depicted and analyzed in detail. In Section 3, we identify the uniform growth state and present the incremental theory for growing bilayered tubular tissues. A linear bifurcation analysis by means of the Stroh formulation and the surface impedance matrix method is carried out to determine the critical growth factor and the correlated wavenumber in Section 4. A comparison of the critical pattern with experimental results is also exhibited. In Section 5, we further compare the post-buckling evolutions as well as the corresponding pattern transitions between experiments and finite element simulations. Finally, some concluding remarks are given in Section 6.

\section{Swelling experiments}
To produce simulacrums of bilayered tubular tissues, we designed a tubular mold consisting of Poly tetra fluoroethylene (PTFE) where the outer radius and height were prescribed by 20mm and 10mm, respectively. The mold contained a cylindrical core with varied diameters so as to create specimens with different sizes. All fabricated tubes were placed in a container where the radius was fixed by 20mm, and a solvent was injected into the empty space to utilize swelling to mimic constrained growth (see schematic in Fig.~\ref{fig1}(a)). 

\subsection*{Materials and methods}
\begin{figure}[!h]
	\centering\includegraphics[width=6.3in]{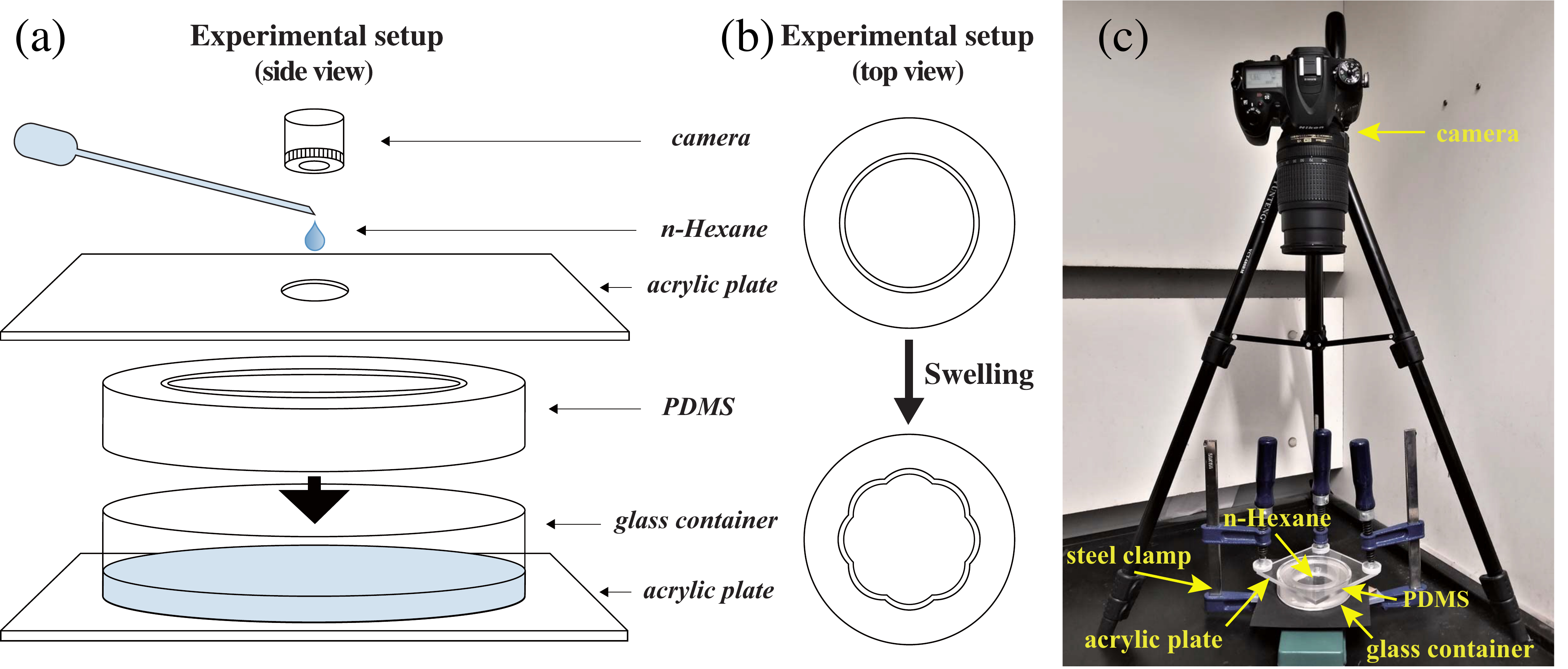}
	\caption{(Color online) Diagram of the experimental platform: (a) side view of all required implements; (b) top view of the deformation process; (c) photo of the experimental setup.}
    \label{fig1}
\end{figure}

Both the inner and outer layers of a bilayered tubular structure were generated with polydimethylsiloxane (PDMS) elastomers (Sylgard 184; Dow Chemical Co., Midland, MI), which have been used as a highly monitored abiotic substitute for biological tissues, such as brain \cite{8}. Then we fabricated bilayered tubular PDMS elastomers following the methodology in \cite{54,55,56}. Since there were two PDMS layers, we first prepared the outer layer, using a prescribed mass ratio of base monomer to curing agent that was mixed and poured over the fabrication mold. The compound was free of air bubbles by virtue of an air pump and then cured in a vacuum desiccator at 100 $^{\circ}$C for 30 minutes. After cooling, the central cylindrical core was replaced by a smaller one to leave space for creating the inner layer. Afterwards, we repeated the previous steps producing the outer layer with a different mixed mass ratio and ultimately acquired a bilayered tubular PDMS structure when carefully peeling it off from the mold. In particular, the elastic moduli of the two layers were varied and the interface between two layers was perfectly bonded during the deformation. Bearing in mind that cured PDMS was practically transparent, we added a little TiO$_2$ powder of white color into the inner layer to distinguish the two layers. It was assumed that the influence of TiO$_2$ on the mechanical property of PDMS can be neglected. The obtained tube was placed in a round glass container where the radius and height were identical to those of the bilayered structure, aiming to restrict growth on the outer boundary and facilitate the attainment of a plane-strain environment. We further used two acrylic plates to confine the axial growth for the purpose of rendering a plane-strain deformation. Note that PDMS will swell when immersed into n-Hexanes. Then, the n-Hexane solution dropped continuously into the glass container through the small circular hole in the center of the acrylic plate such that the specimen was always immersed into n-Hexanes completely during the entire swelling process. In this experimental setup, the n-Hexanes entered into the structure through the inner surface, giving rise to a swelling-induced deformation. As deformation increased, surface instabilities took place, which were recorded by a camera on the top. The detailed experimental setup is shown in Figure \ref{fig1}.

\subsection*{Material parameter characterization}
For convenience, the base monomer is denoted by $S_1$ while the curing agent is signified by $S_2$. As a result, we let $S_1:S_2$ to represent the mass ratio between base monomer and curing agent. It is known that the elastic modulus of PDMS elastomer is dependent on the mixed mass ratio of the two chemical reagents (base monomer and curing agent) \cite{55}, we hence employed that the mass ratio $S_1:S_2$ was $3:1$, $10:1$, $25:1$, respectively. 

It is well understood that the ratio of the shear modulus of the film to that of the substrate in planar and curved film-substrate structures is critical in determining the bifurcation nature \cite{34,57,58}, wrinkled pattern \cite{33,59,60}, and morphological evolution \cite{61,62,63,64}. Specifically, The mechanical properties of PDMS are known to vary with the cross-linked temperature as well as the associated duration time \cite{65}. So it is important to determine the relative stiffnesses of the two layers in the bilayered PDMS structure. For that purpose, we performed a material characterization test to determine the material constant in a specified constitutive model. Furthermore, two distinct experiments were conducted, including a uni-axial tensile test and a pure shear one.

Since PDMS is almost incompressible, we then employ the incompressible neo-Hookean model as the material constitution of PDMS elastomer, and the strain energy function $W$ is given by 
\begin{equation}
	W=\dfrac{1}{2}\mu(\lambda_1^2+\lambda_2^2+\lambda_3^2-3),
	\label{eq1}
\end{equation}
where $\mu$ denotes the shear modulus and $\lambda_i$ $(i=1,2,3)$ stands for the principal stretch. Then the aim is to identify the values of $\mu$ for different specimens.  

\begin{figure}[!h]
    \centering
    \includegraphics[width=5in]{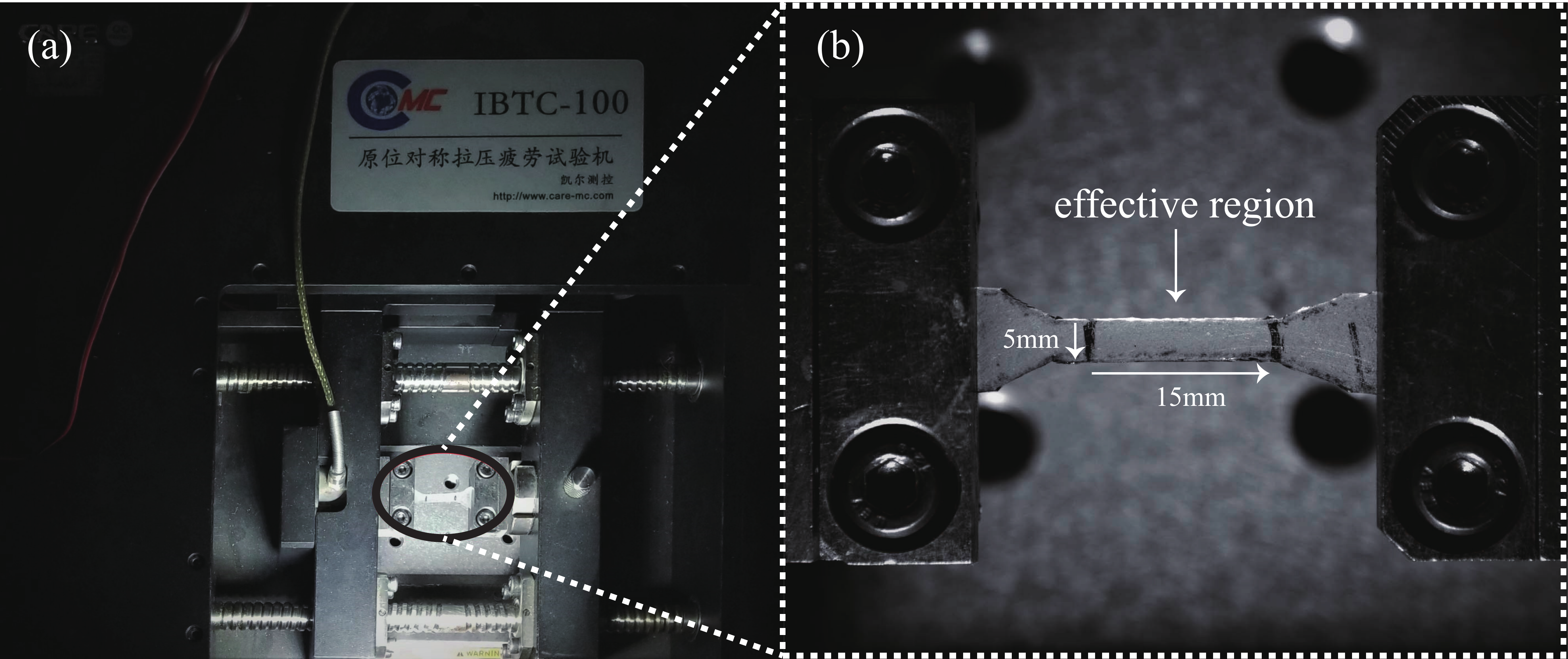}
    \caption{Sketch of the uni-axial tensile test: (a) the experimental system; (b) blow-up of the sample with the corresponding size of the effective region.}
    \label{fig2}
\end{figure}

In light of the clamping methodology in Zhang et al. \cite{66}, uni-axial tensile tests were carried out using tensile test system (modal: CARE Measurement $\&$ Control, IBTC-100) where the relationship between force and displacement can be automatically documented. Figure \ref{fig2} displays an experimental setup together with a dumb-bell specimen. It can be seen that the original fixture of the test system was used. The effective size for strain measurement enclosed within the two black lines in the dumb bell shaped sample was 15 mm $\times$ 5 mm. The thicknesses were given by 1.74 mm, 0.88 mm, and 0.75 mm for $S_1:S_2=3:1$, $10:1$, and $25:1$, respectively. Referring to the comparative studies shown in \cite{67,68,69}, we calibrated the material parameter in the neo-Hookean model to the experimental data according to the method of least squares. For a uni-axial stretch, the Cartesian rectangular coordinates system is used, and the nominal stress tensor reads

\begin{equation}
	\mathbf{S}=\frac{\partial W}{\partial \mathbf{F}}-p\mathbf{F}^{-1},
	\label{eq2}
\end{equation}
where $p$ stands for the Lagrange multiplier enforcing the incompressibility condition, and the deformation gradient $\mathbf{F}$ for uni-axial extension writes
\begin{equation}
	\mathbf{F}=\lambda_1 \bm e_1\otimes\bm e_1+\lambda_2\bm e_2\otimes\bm e_2+\lambda_3\bm e_3\otimes\bm e_3.
	\label{eq3}
\end{equation}
In the above formula, we have denoted the common orthonormal basis in both the reference (initial) and current (stretched) configurations by $\{{\bm e}_1, {\bm e}_2, {\bm e}_3 \}$.

It can be readily deduced from the traction-free conditions in the $x_2$- and $x_3$-directions that $\lambda_2=\lambda_3=\lambda_1^{-1/2}$ and $p=\mu/\lambda_1$. Furthermore, in virtue of the loading condition and the method of least squares, we arrive at
\begin{equation}
	\mu=\dfrac{\sum_{k=1}^m s_k( \lambda_{1k}-1/\lambda_{1k}^2)}{\sum_{k=1}^m ( \lambda_{1k}-1/\lambda_{1k}^2)^2},
	\label{eq2_5}
\end{equation}
where $\lambda_{1k}$ indicates the stretch in the $x_1$-direction and $s_k$ the associated nominal stress measured in experiments for the $k$th test point while $m$ denotes the total number of test points. 

\begin{figure}[!h]
	\centering
	\subfigure[$S_1:S_2=3:1$, $\mu=0.5439$ MPa.]{\includegraphics[scale=0.59]{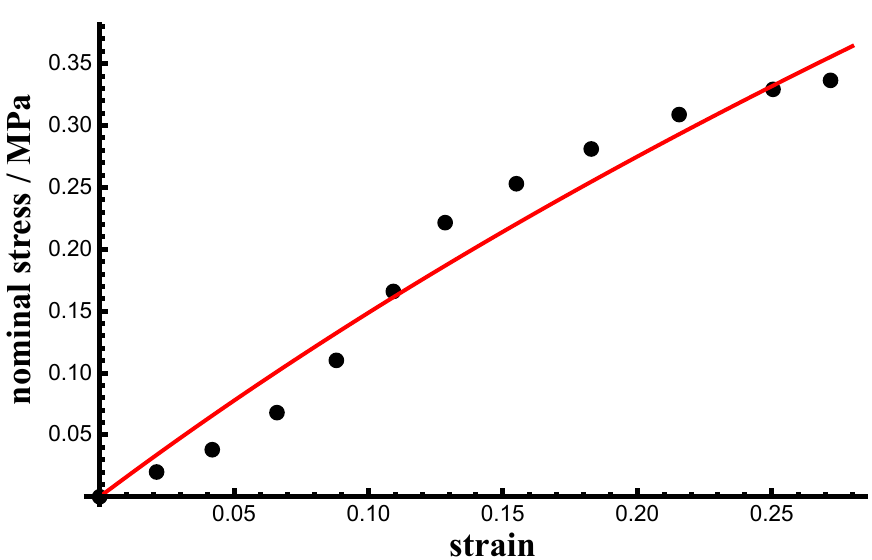}{\label{fig3a}}}\hspace{3mm}
\subfigure[$S_1:S_2=10:1$, $\mu=0.7686$ MPa.]{\includegraphics[scale=0.59]{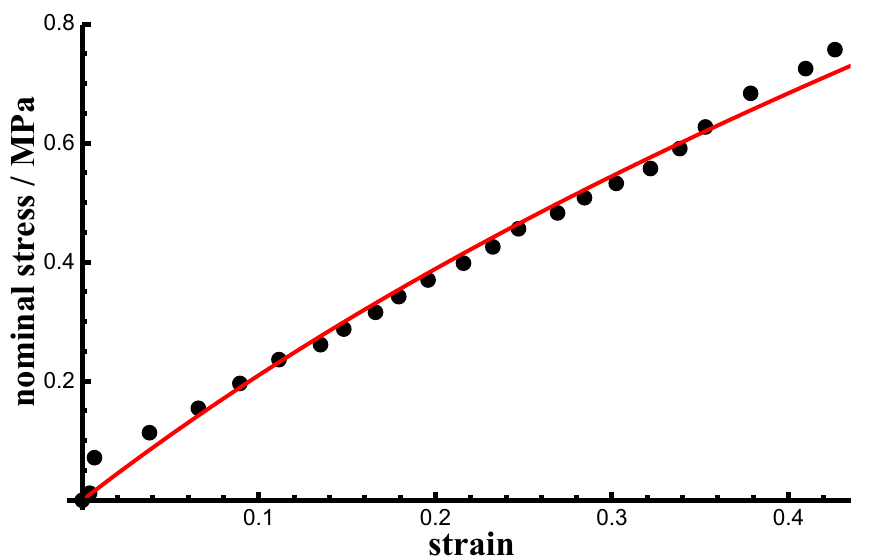}{\label{fig3b}}}\hspace{3mm}
\subfigure[$S_1:S_2=25:1$, $\mu=0.1228$ MPa.]{\includegraphics[scale=0.59]{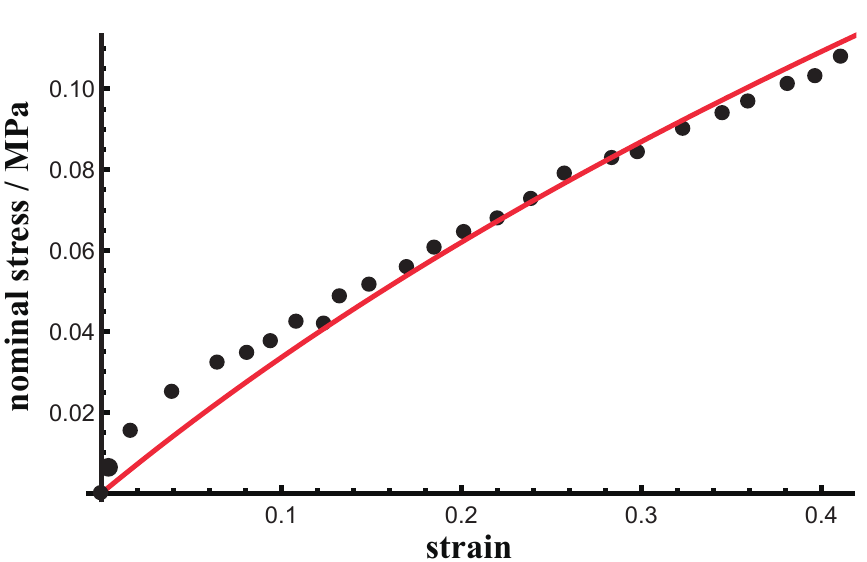}{\label{fig3c}}}
	\caption{(Color online) Fitting curves (red lines) and the experimental data (black dots) for different mass ratios in uni-axial tensile tests. The strain is given by $\lambda_1-1$. }
	\label{fig3}
\end{figure}

The fitting curves for three typical mass ratios are drawn in Figure \ref{fig3} where the red lines imply the theoretical predictions and the black dots correspond to experimental data. In particular, the evaluated material constants are presented below each subfigure. It can be seen that the shear modulus $\mu$ in the neo-Hookean model is no longer monotonically dependent on the mass ratio of the two chemical reagents in our prepared samples.

\begin{figure}[!h]
	\centering\includegraphics[width=5in]{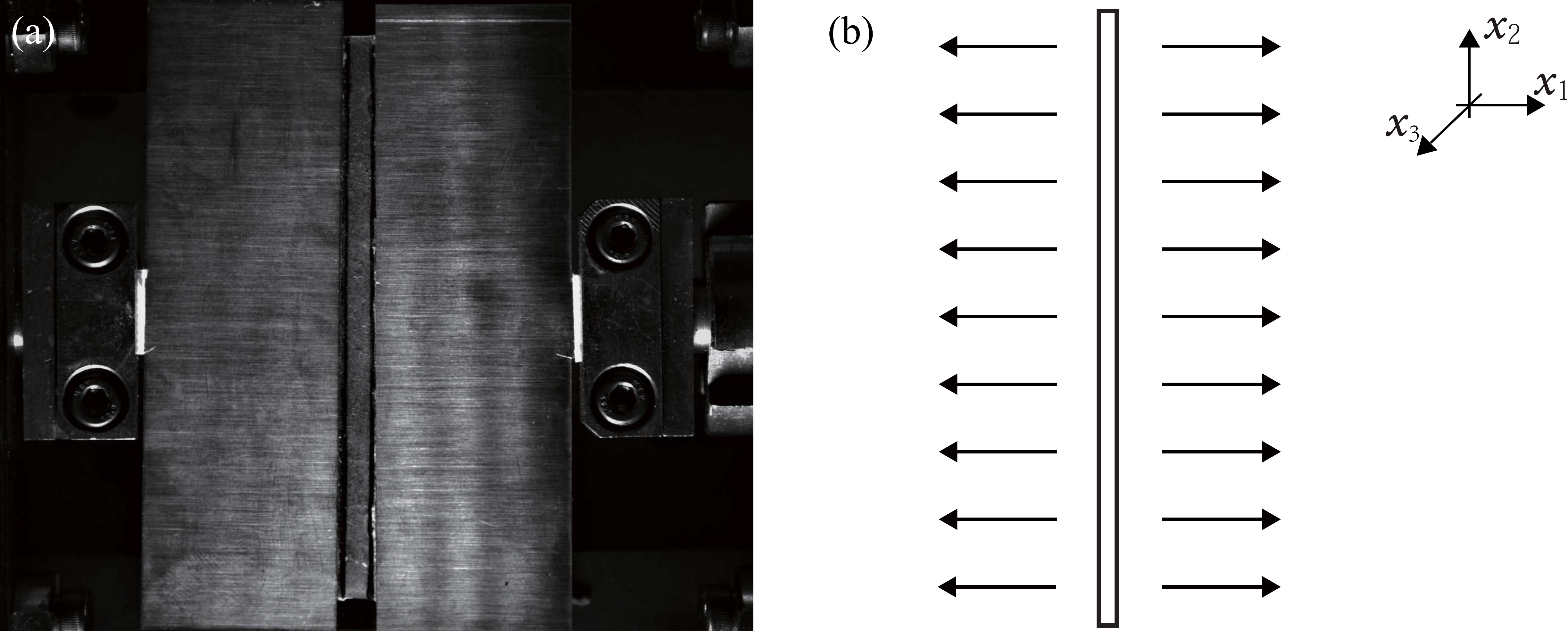}
	\caption{Sketch of the pure shear test: (a) the $in$ $situ$ state in an experiment; (b) diagram of the sample and the arrows indicate the direction of extension.}
	\label{fig4}
\end{figure}

\begin{figure}[!h]
	\centering
	\subfigure[$S_1:S_2=3:1$, $\mu=0.4902$ MPa.]{\includegraphics[scale=0.8]{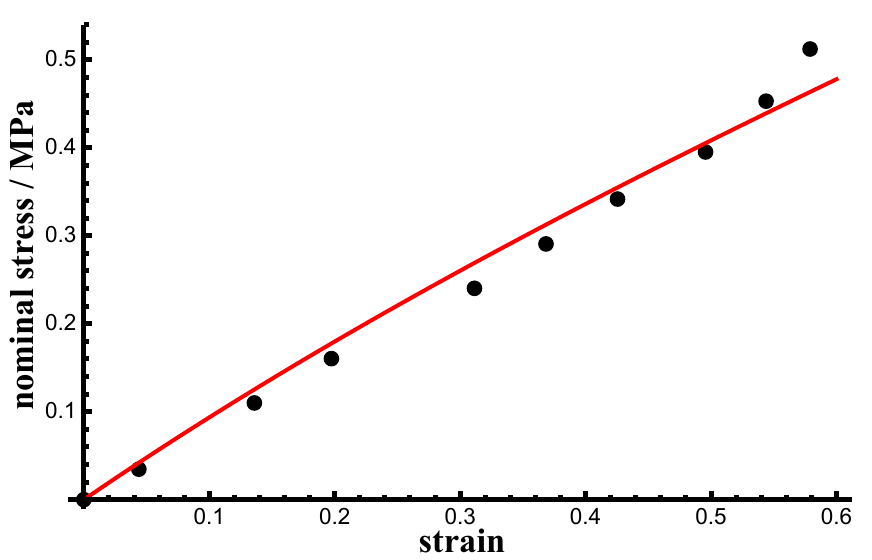}{\label{fig5a}}}\hspace{5mm}
	\subfigure[$S_1:S_2=10:1$, $\mu=0.791$ MPa.]{\includegraphics[scale=0.8]{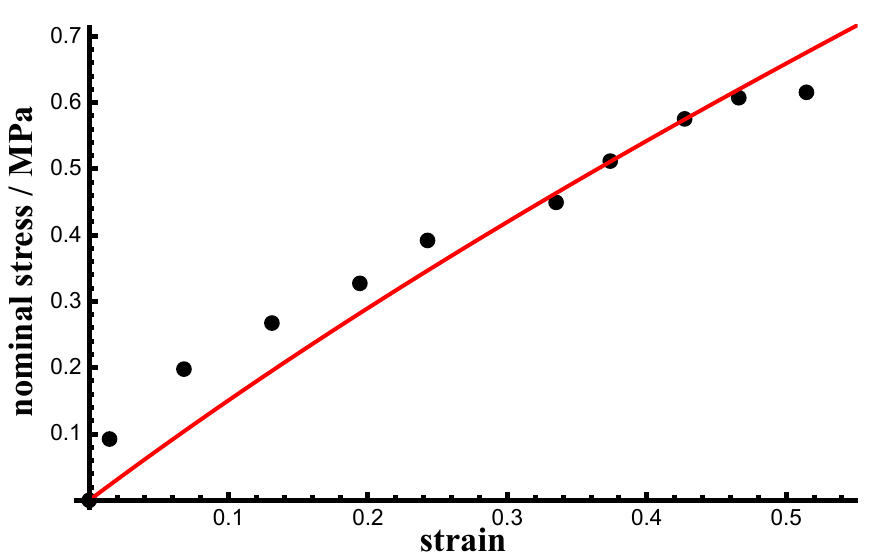}{\label{fig5b}}}	\caption{(Color online) Fitting curves (red lines) and the experimental data (black dots) for different mass ratios in pure shear tests. The strain is determined by $\lambda_1-1$.}
	\label{fig5}
\end{figure}

We further conducted a pure shear experiment in order to verify the above results. Details of pure shear deformation can be found in \cite{70,71}. Figure \ref{fig4} shows the experimental setup as well as the shape of a sample. It should be pointed out that the sample for pure shear test usually possesses an extremely low aspect-ratio. From the previous experiments, we find that the shear modulus $\mu$ is quite small when $S_1:S_2=25:1$. In fact, a softer PDMS becomes sticky. Therefore, it is difficult to produce a proper sample in this case since the peeling procedure would destroy the narrow structure and we abandon this test for $S_1:S_2=25:1$. Seen from Figure \ref{fig4}(a), the specimen was first glued to a wide iron sheet and then clamped to the fixture in the test machine. In a pure shear deformation, the second principal stretch is prescribed by unity such that $\lambda_2=1$ and $\lambda_3=\lambda_1^{-1}$ in equation (\ref{eq3}). The width to length ratio was specified by $1/20$, and the shear modulus can be calibrated by the same solution strategy as that in the uni-axial tensile test. For brevity, we omit the technique details and immediately plot the results in Figure \ref{fig5} for two different mass ratios of the base monomer to the curing agent, where the corresponding material constant is shown below each subfigure. It is found that the results based on the pure shear tests are close to those on the uni-axial tensile trails, and the relative errors are less than 10$\%$. Consequently, we take an averaged value for the shear modulus when $S_1:S_2$ equals to $3:1$ and $10:1$. For the other scenarios, we employ the prediction by the uni-axial tensile experiment. Finally, we summarize the results of material characterization in Table \ref{table1}.

\begin{table}[h]
   \begin{center}
    \caption{The experimentally fitted shear moduli for different mass ratios.}
    \bigskip
    \begin{tabular}{cccc}
     \hline $S_1:S_2$ & $3:1$ & $10:1$ & $25:1$\\
     \hline shear modulus (MPa) &0.5171 &0.7798 &0.1228\\  \hline
    \end{tabular} \label{table1}
    \end{center}
\end{table}

Currently, we have determined the material constant in the incompressible neo-Hookean model and have listed the values in Table \ref{table1}. It turns out that the fitted shear moduli obtained by different material characterization tests are almost identical. We emphasize that these values of $\mu$ will be employed as the virtual material parameters in the theoretical analysis as well as the finite element simulation for exhaustive comparisons of the pattern formation and transition. Additionally, we point out that the calibrated data in Table \ref{table1} deviate from the counterparts given in \cite{72} relying on a compression setting, especially for the case of $3:1$. The divergence is chiefly caused by the fact that the curing temperature and time are both dissimilar. We further stress again that the material characterization test merely aims to give the elastic modulus of the samples used in our experiment, and the corresponding results are not expected to shed light on the relationship between elastic modulus and mass ratio for PDMS.

\subsection*{Experimental results}
We prepared in total five specimens with different geometries and modulus ratios for swelling experiment (abbreviated as $sp_i$ corresponding to $i$th specimen) on the basis of the tubular mold and the accurate data are exhibited in Table \ref{table2}. In particular, the $sp_1$ ($sp_3$) and $sp_2$ ($sp_4$) samples share the same modulus ratio but occupy different thickness of the inner layer. The $sp_1$ ($sp_2$) and $sp_3$ ($sp_4$) samples are of the same geometry but of various modulus ratios. Moreover, we intend to utilize the last specimen $sp_5$ to view the bifurcation behavior when the shear modulus of the inner layer is marginally greater than that of the outer layer. Actually, this setting implies that we are concerned with the influence of modulus ratio and the thickness of inner layer on pattern formation in growing tubular tissues.

\begin{table}[!ht]
   \begin{center}
    \caption{The detailed information of five samples used in swelling experiment.}
    \resizebox{\textwidth}{15mm}{
    \begin{tabular}{ccccccc}
     \hline label&inner radius (mm)&interfacial radius (mm)&outer radius (mm)&$S_1:S_2$ (inner layer)&$S_1:S_2$ (outer layer)&modulus ratio $\approx$\\
    \hline $sp_1$&26.8&28&40&$3:1$&$25:1$&4.21\\
   $sp_2$&27.6&28&40&$3:1$&$25:1$&4.21\\
   $sp_3$&26.8&28&40&$10:1$&$25:1$&6.35\\    
   $sp_4$&27.6&28&40&$10:1$&$25:1$&6.35\\
   $sp_5$&26.8&28&40&$11:1$&$10:1$&1\\  \hline
    \end{tabular}} \label{table2}
    \end{center}
\end{table}

We carried out swelling trials for these samples in the experimental platform illustrated in Figure \ref{fig1}. Since PDMS absorbs n-Hexanes, we shall constantly supplement the corresponding solvent through the circular whole in the top plate. Due to the restrictions in both the top and bottom surfaces as well as the outer boundary, a plane strain deformation mimicking constrained growth in the circumferential plane occurred. A swelling process may take several hours to trigger surface instabilities and further to create pattern transition. In the subsequent analysis, we shall present the experimental results in detail.

Before preceding further, we define some dimensionless parameters in order to facilitate comparisons of the experimental and theoretical predictions. We let $A$, $B$, and $C$ to represent the inner radius, interfacial radius, and outer radius, respectively, and write two dimensionless geometrical parameters $A^*=A/C$ and $B^*=B/C$. In addition, the dimensionless parameter $\beta$ depicts the ratio of the shear modulus of the inner layer to that of the outer layer.

\begin{figure}[!h]
	\centering
	\subfigure[Primary deformation.]
	{\includegraphics[scale=0.2]{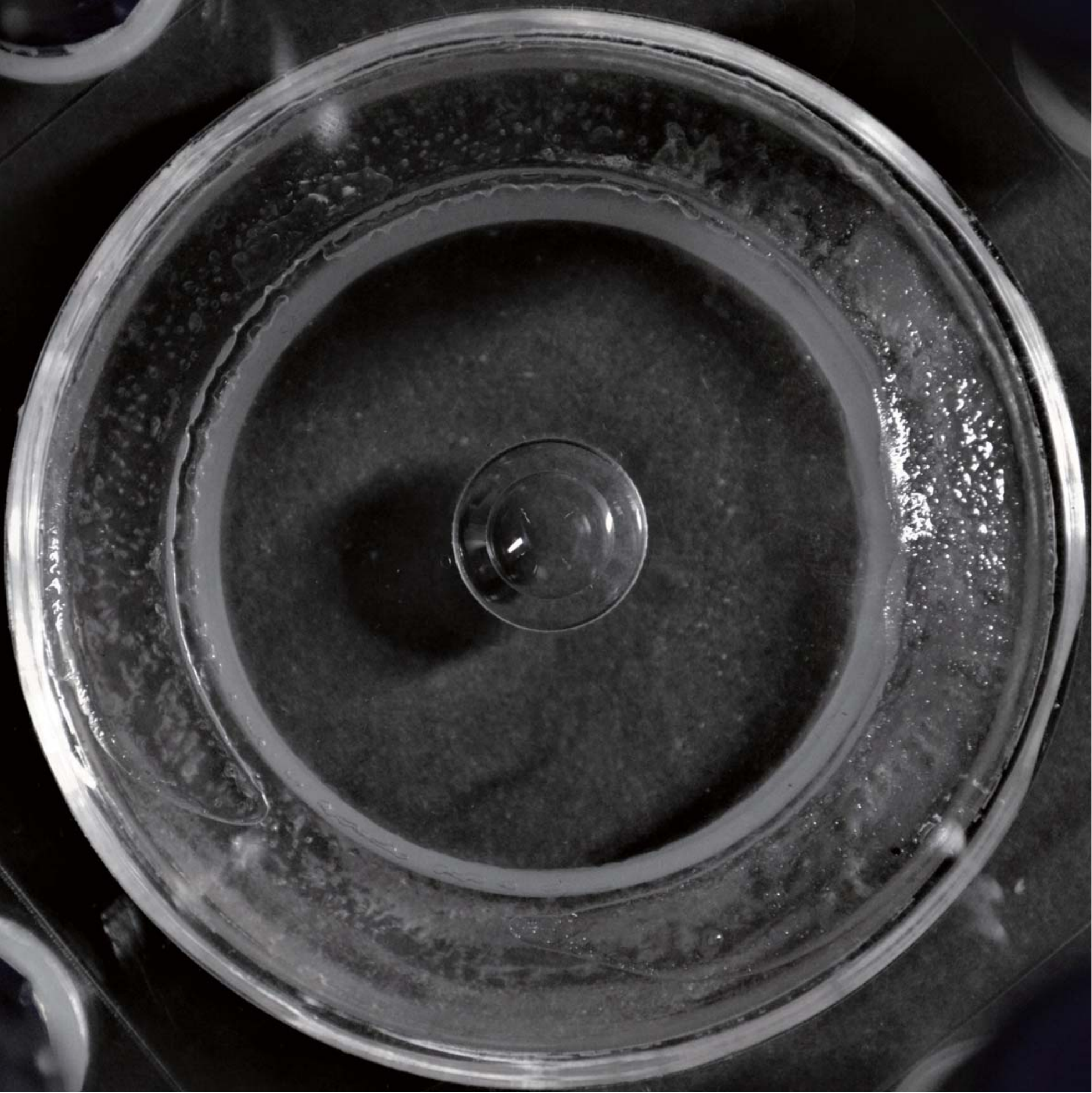}{\label{fig6a}}}
	\subfigure[Wrinkled strate.]
         {\includegraphics[scale=0.2]{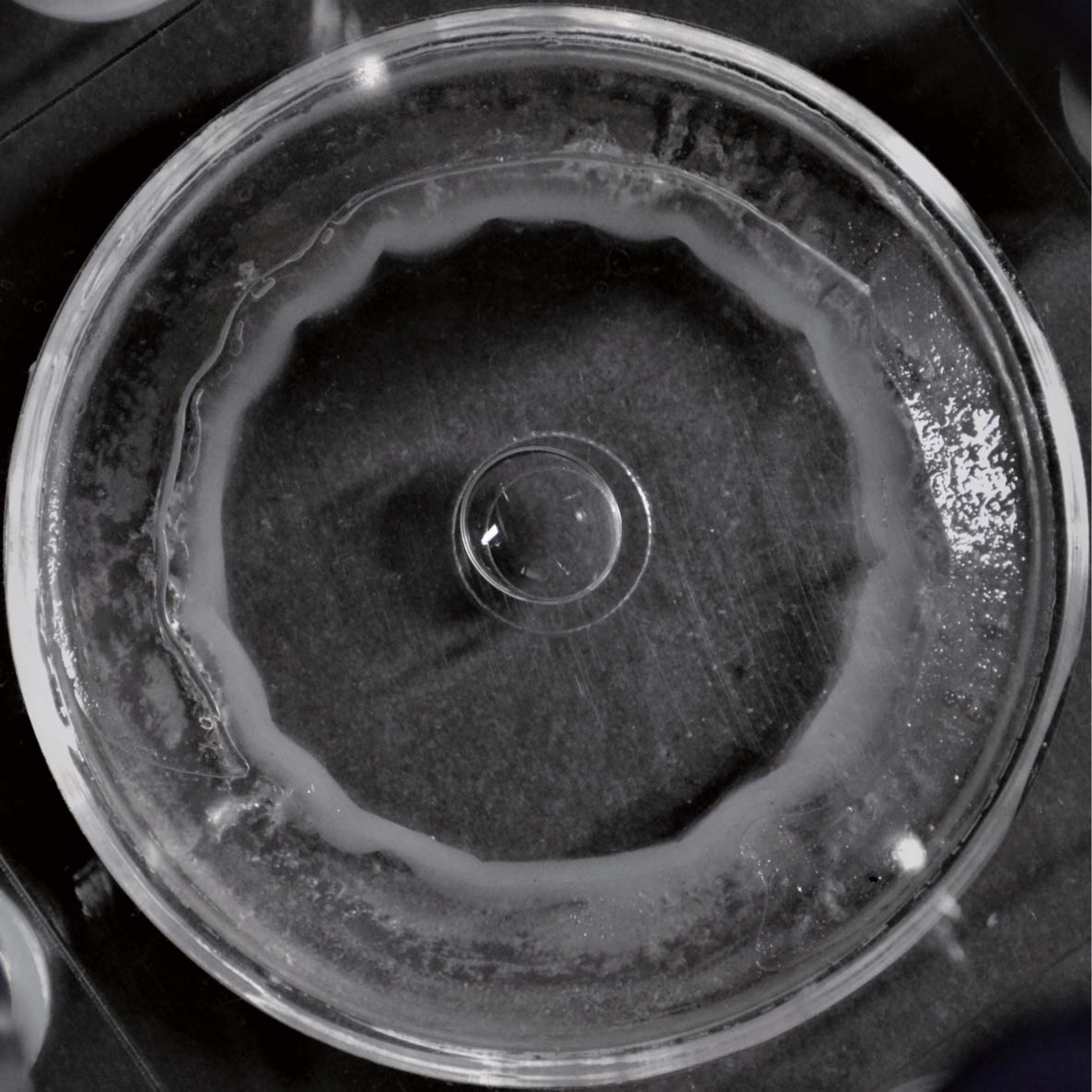}{\label{fig6b}}}
	\subfigure[Period-doubling with crease.]
	{\includegraphics[scale=0.2]{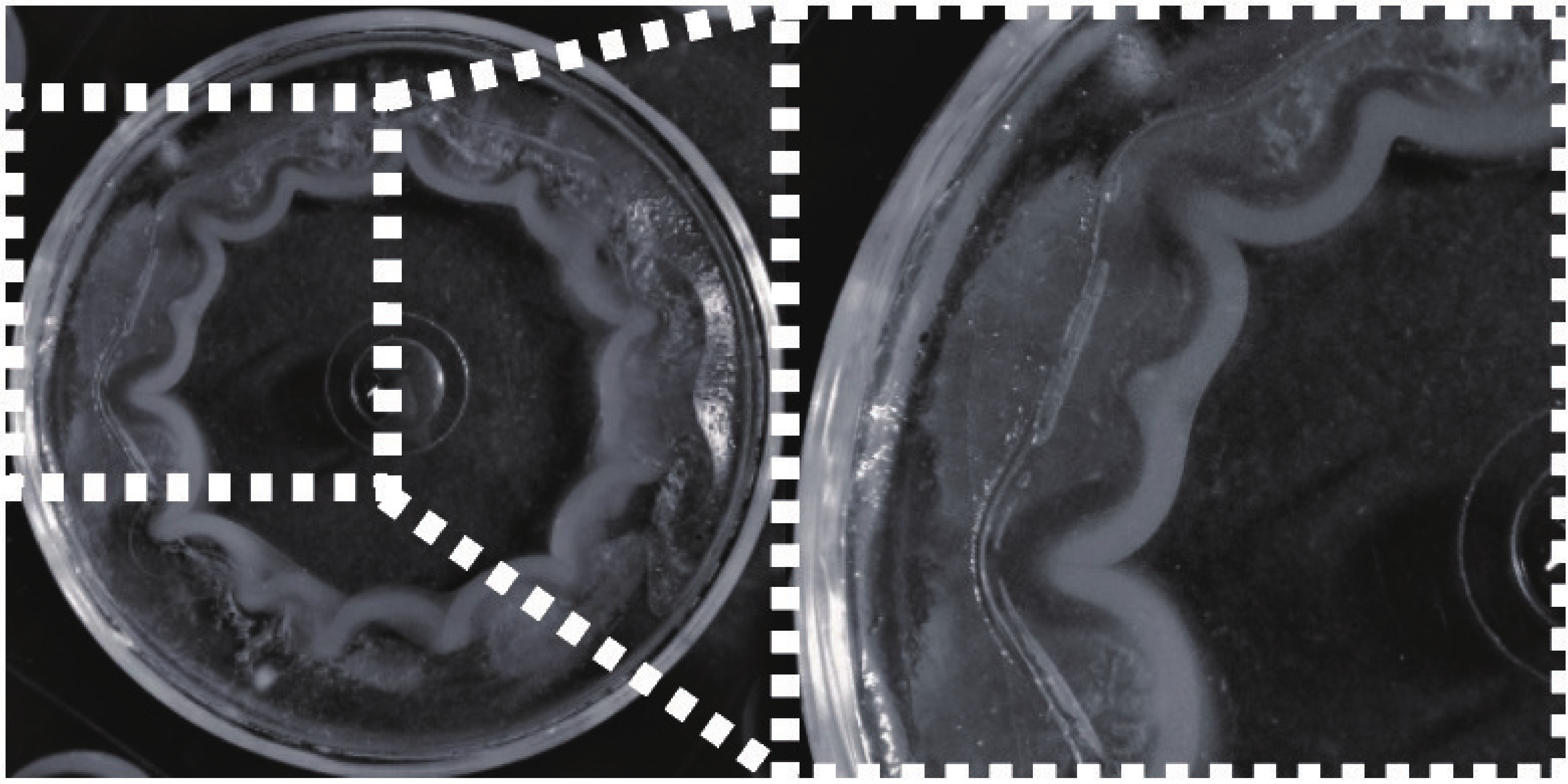}{\label{fig6c}}}
	\caption{Swelling-induced surface instabilities for sample $sp_1$. The dimensionless parameters are identified from Table \ref{table2} by $A^*=0.67$, $B^*=0.7$, and $\beta\approx4.21$. The first bifurcation directed a wavy pattern where the wavenumber was 14, and the rightmost subfigure is a blowup of the part enclosed in the dashed rectangle.}
	\label{fig6}
\end{figure}

Figure \ref{fig6} plots the three typical deformed configurations of the swelling-induced deformation, i.e. a primary state, a wavy pattern, and a period-doubling morphology with co-existing wrinkles and creases. The normalized parameters were $A^*=0.67$, $B^*=0.7$, and $\beta\approx4.21$. Firstly, n-Hexanes enter into PDMS structure from the inner surface and generate circumferential compressive stress, which incurs a sinusoidal profile after it passes a critical value, as illustrated in Figure \ref{fig6b}. There were in total 14 waves and the amplitude of wrinkles gradually increases as dilatation continues. Finally, a specific surface pattern emerged where a period-doubling mode can be observed, as shown in the left subfigure in Figure \ref{fig6c}. Particularly, unlike an ordinary period-doubling secondary bifurcation where no self-contact appears, each period is composed of a wrinkle and a crease, as displayed in the right subfigure in Figure \ref{fig6c}. Note that this special mode has  been studied by Liu et al. \cite{35} in a growing tubular tissue with exponentially decayed shear modulus and by Chen et al. \cite{73} in a compressive graded half-space where the Young's modulus declines exponentially. It was shown by Fu and Cai \cite{61} that period-doubling secondary bifurcation (without self-contact) may take place at a critical compressive strain in film-substrate structures only if when $\beta$ is nearly greater than 5.8. Although this vital value may be slightly varied in curved systems, this critical condition $\beta\approx4.21<5.8$ can still be used to unravel underlying mechanism behind the observed pattern transition. 

\begin{figure}[!h]
	\centering
	\subfigure[Primary deformation.]
	{\includegraphics[scale=0.25]{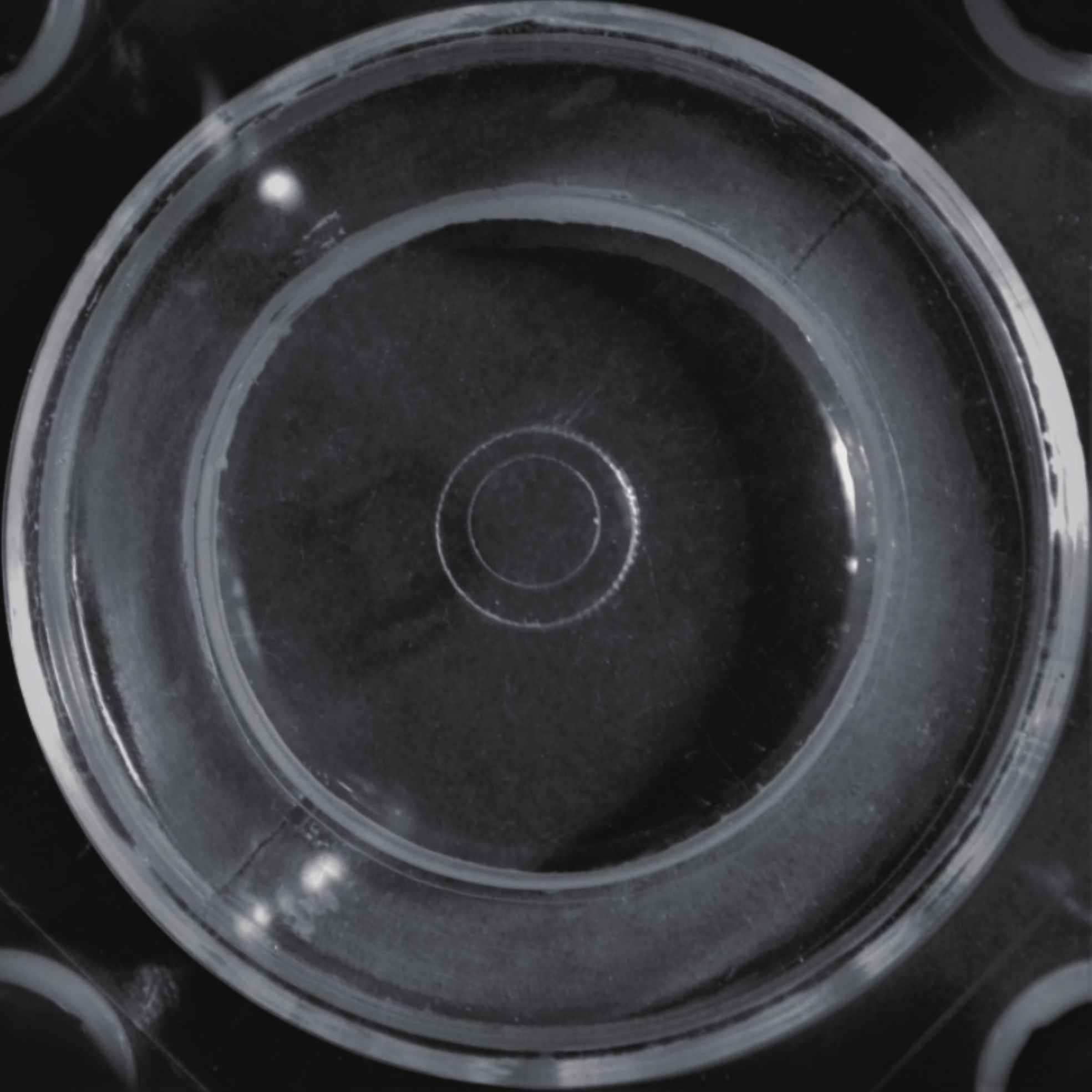}{\label{fig7a}}}
	\hspace{2.5mm}
	\subfigure[Interfacial wrinkling.]
         {\includegraphics[scale=0.25]{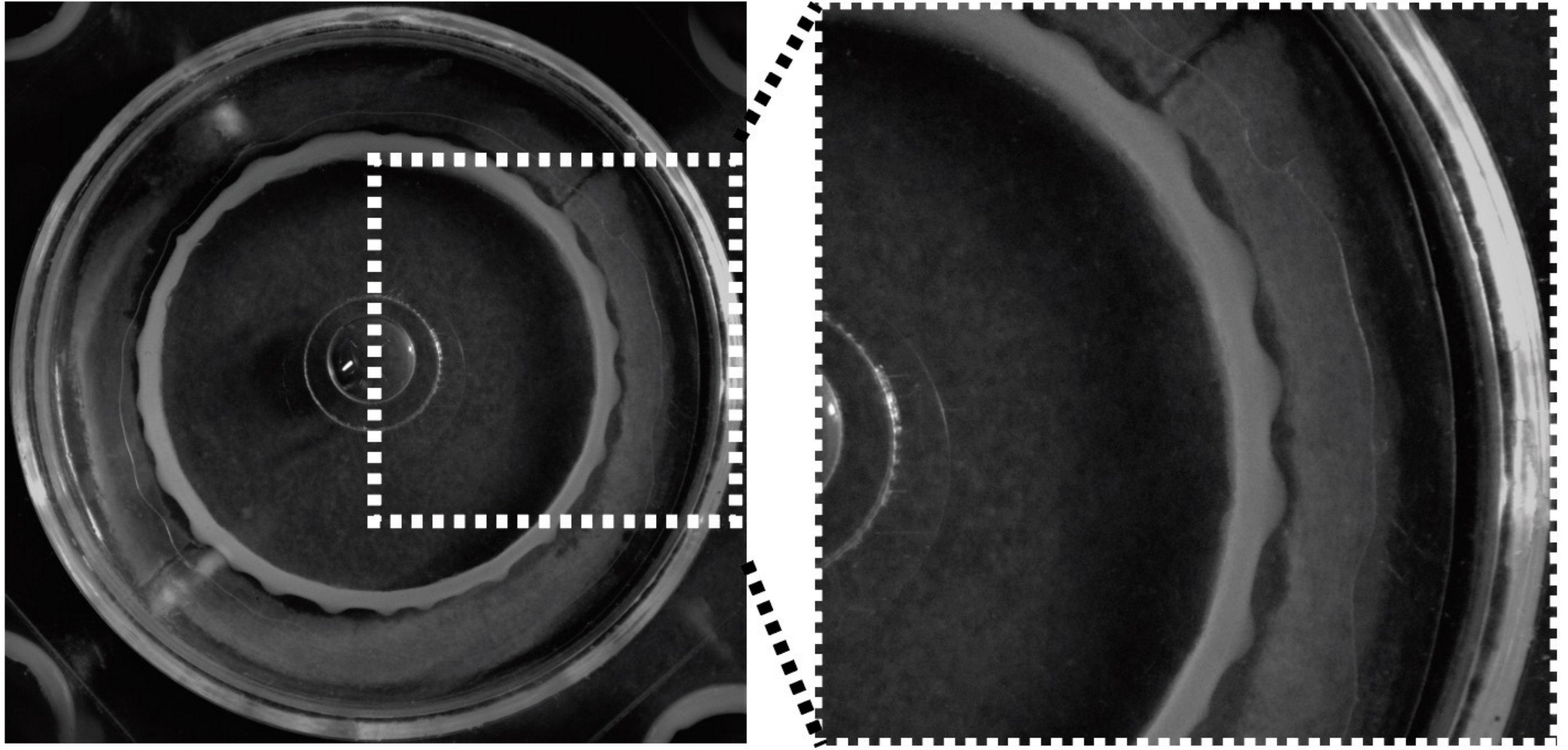}{\label{fig7b}}}
	\caption{Swelling-induced instabilities for sample $sp_2$. The dimensionless parameters are identified from Table \ref{table2} by $A^*=0.69$, $B^*=0.7$, and $\beta\approx4.21$. The first bifurcation created a wavy pattern on the interface containing nearly 28 wrinkles, and the rightmost subfigure is a blowup of the part enclosed in the dashed rectangle.}
	\label{fig7}
\end{figure}

Compared with $sp_1$, only the thickness of the inner layer was changed for sample $sp_2$. Figure \ref{fig7} exhibits the deformations in distinct stages. In contrast to $sp_1$, the experimental result for $sp_2$, sketched in Figure \ref{fig7}, unfolds a diverse phenomenon. Remarkably, neither a surface wrinkling nor a transition between different patterns was witnessed in the swelling process. After a careful check, we find that the wrinkled morphology originated near the interface between two layers and no other morphology was discovered in our experiment until the deformation was almost suspended. Besides, the inner surface remained almost flat in the deformation. We suspect that a large growth factor may be required to trigger a pattern transition for a soft tissue of the same parameters as those of the sample $sp_2$. However, with swelling process proceeding long enough, the osmotic pressure that drives the deformation between PDMS and n-Hexane tended to be zero, and swelling deformation ceased before other modes appear. Generally speaking, interfacial creases occurred in a two-hydrogel system suffering confined swelling \cite{74}. In growing bilayered tubes, Razavi et al. \cite{75} uncovered that interfacial creases is preferred compared to surface creases if the outer layer is very thin. However, interfacial wrinkles have not been reported in tubular structures as far as the authors' knowledge. Nevertheless, this interesting instability is beyond the scope of this study, and we leave it for a further investigation.

\begin{figure}[!h]
	\centering
	\subfigure[Primary deformation.]
	{\includegraphics[scale=0.2]{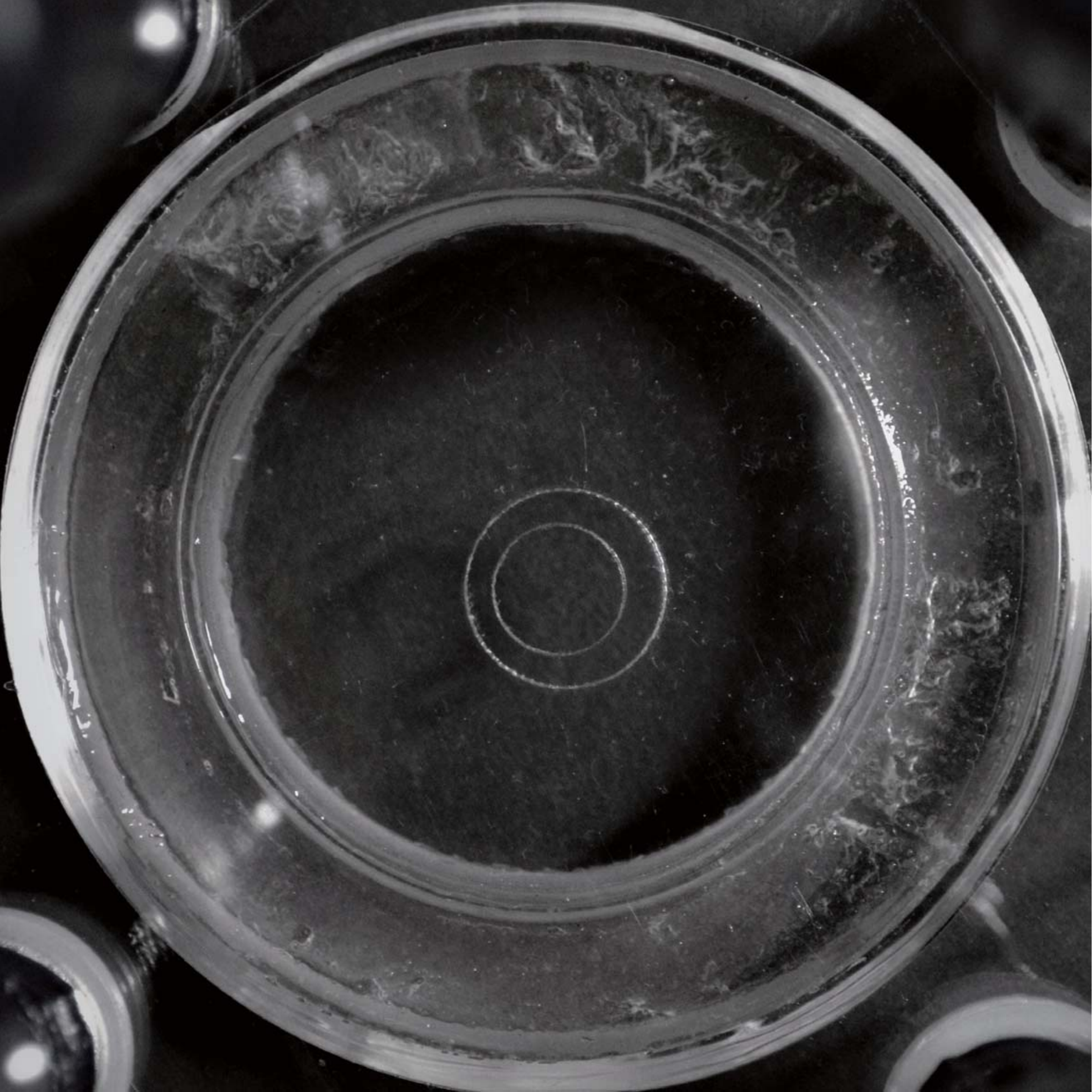}{\label{fig8a}}}
	\subfigure[Sinusoidal wrinkles.]
         {\includegraphics[scale=0.2]{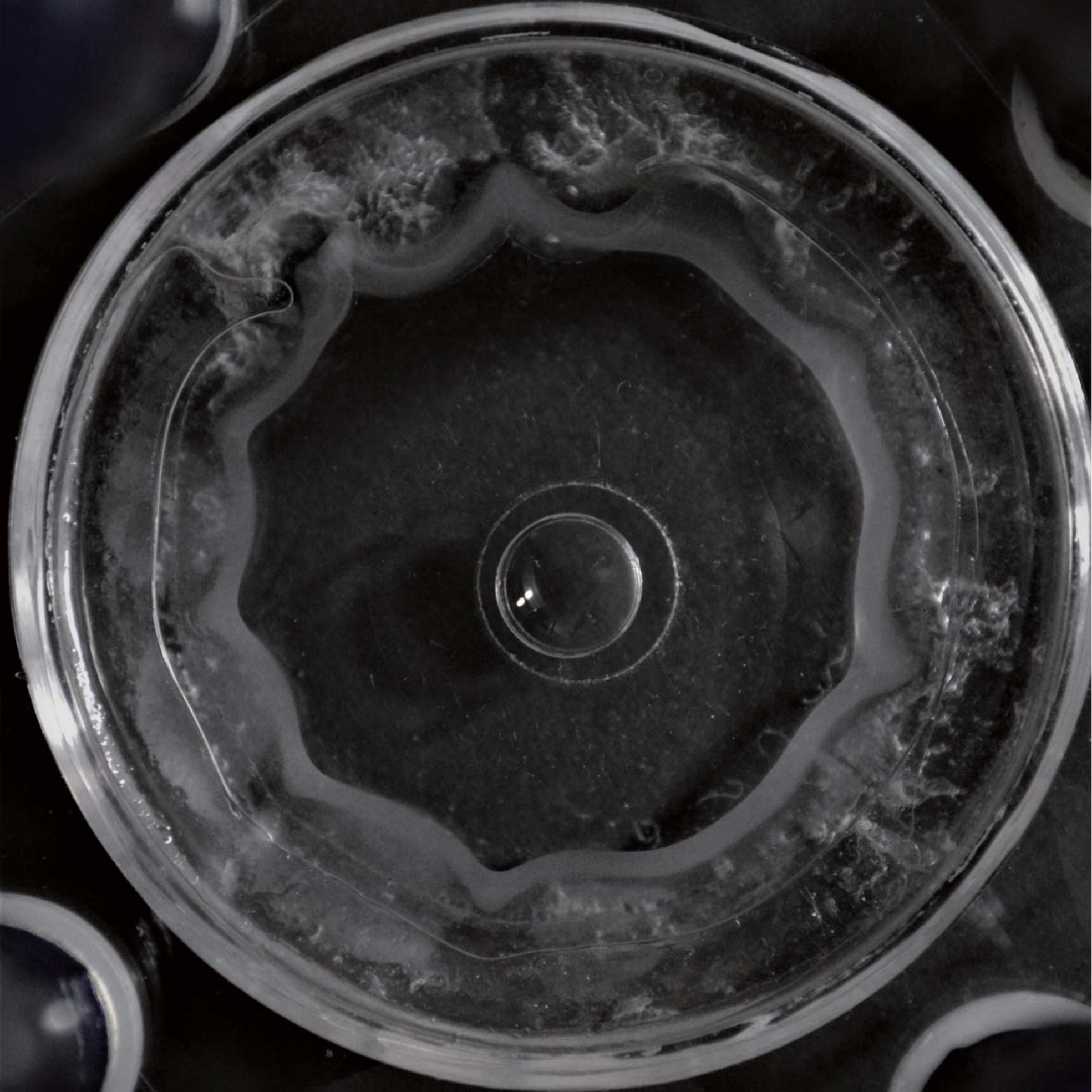}{\label{fig8b}}}
	\subfigure[Initiation of crease.]
	{\includegraphics[scale=0.2]{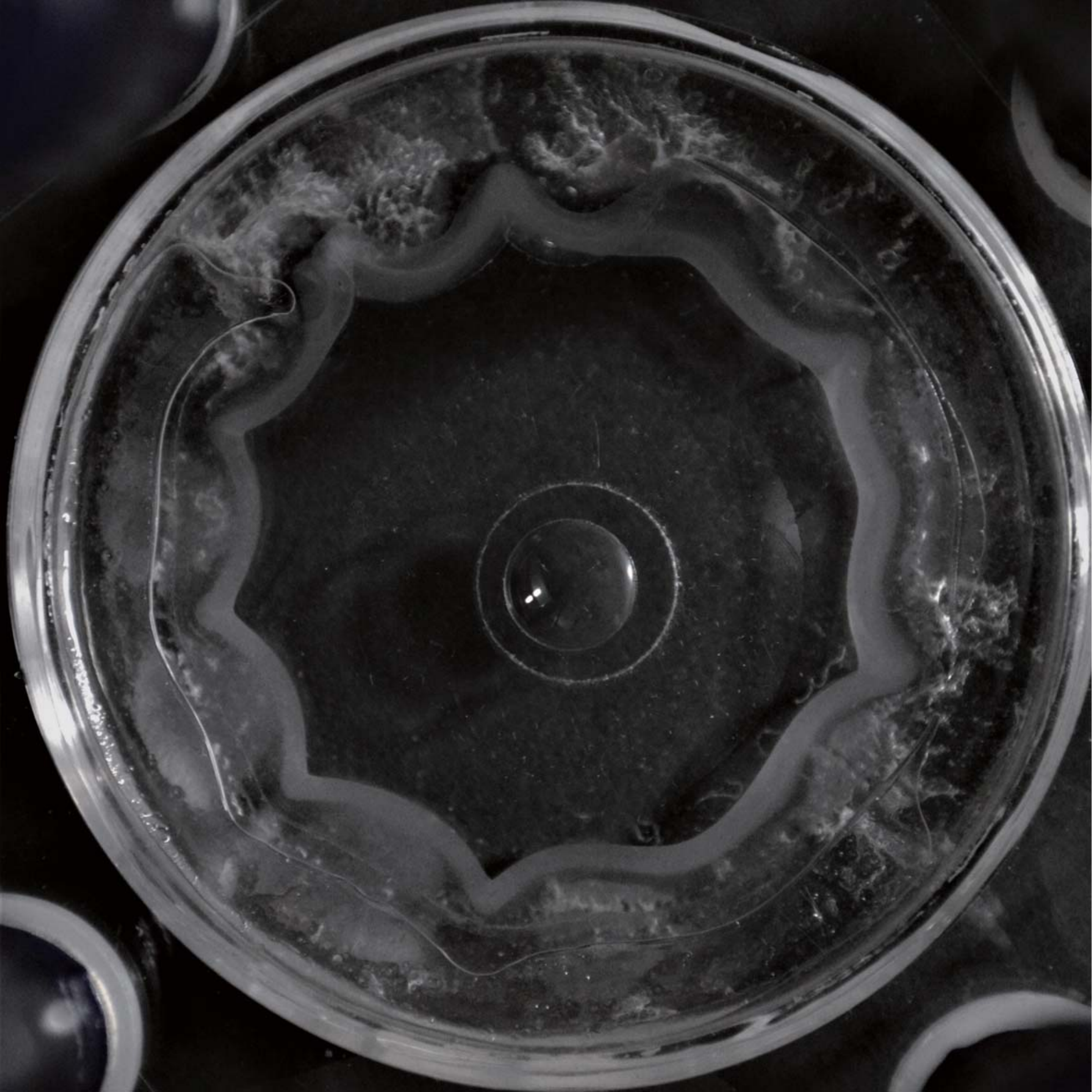}{\label{fig8c}}}
	\subfigure[Deep creases.]
	{\includegraphics[scale=0.2]{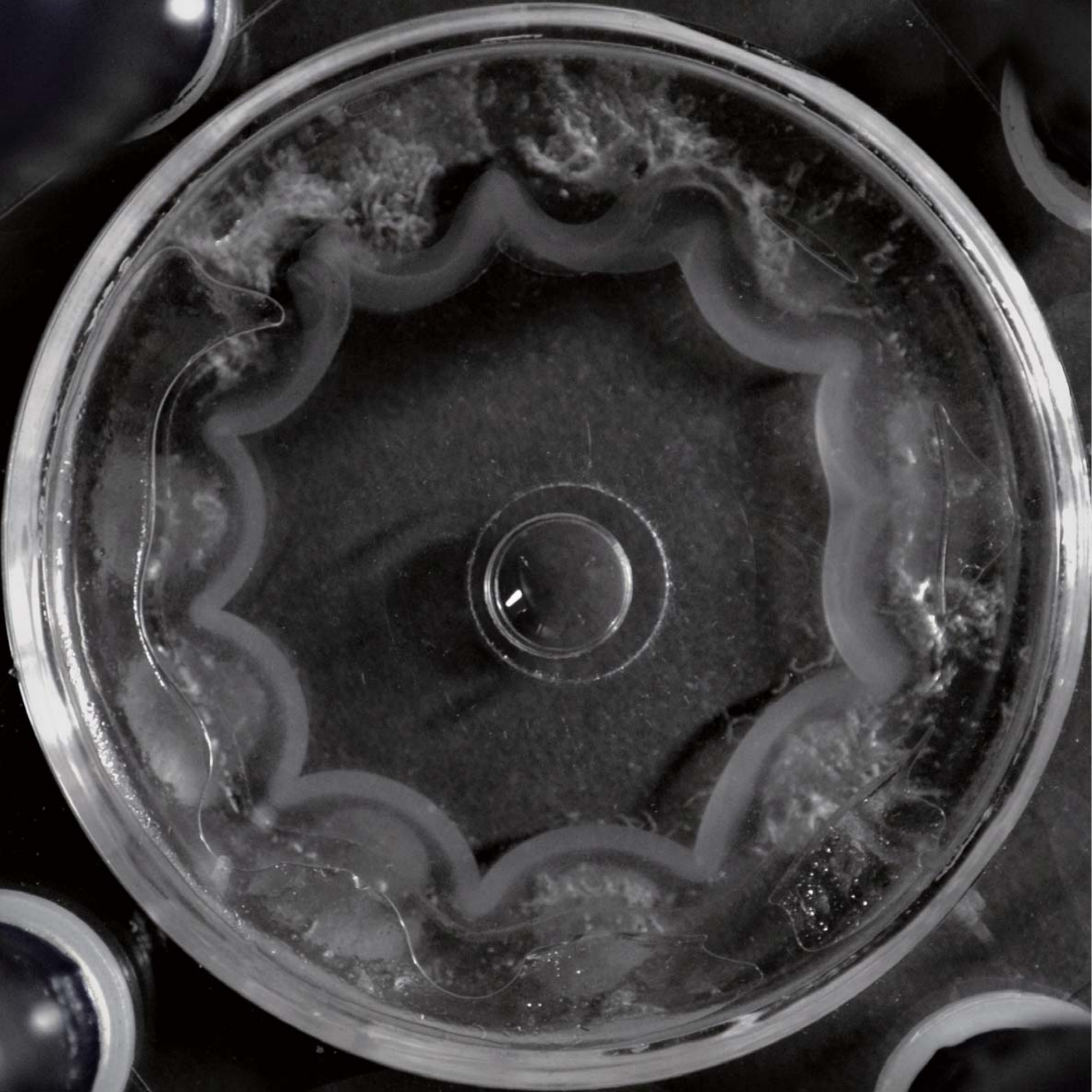}{\label{fig8d}}}
	\caption{Swelling-induced surface instabilities for sample $sp_3$. The dimensionless parameters are identified from Table \ref{table2} by $A^*=0.67$, $B^*=0.7$, and $\beta\approx6.35$. The first bifurcation created a wavy pattern containing 11 wrinkles.}
	\label{fig8}
\end{figure}

\begin{figure}[!h]
	\centering
	\subfigure[Primary deformation.]
	{\includegraphics[scale=0.2]{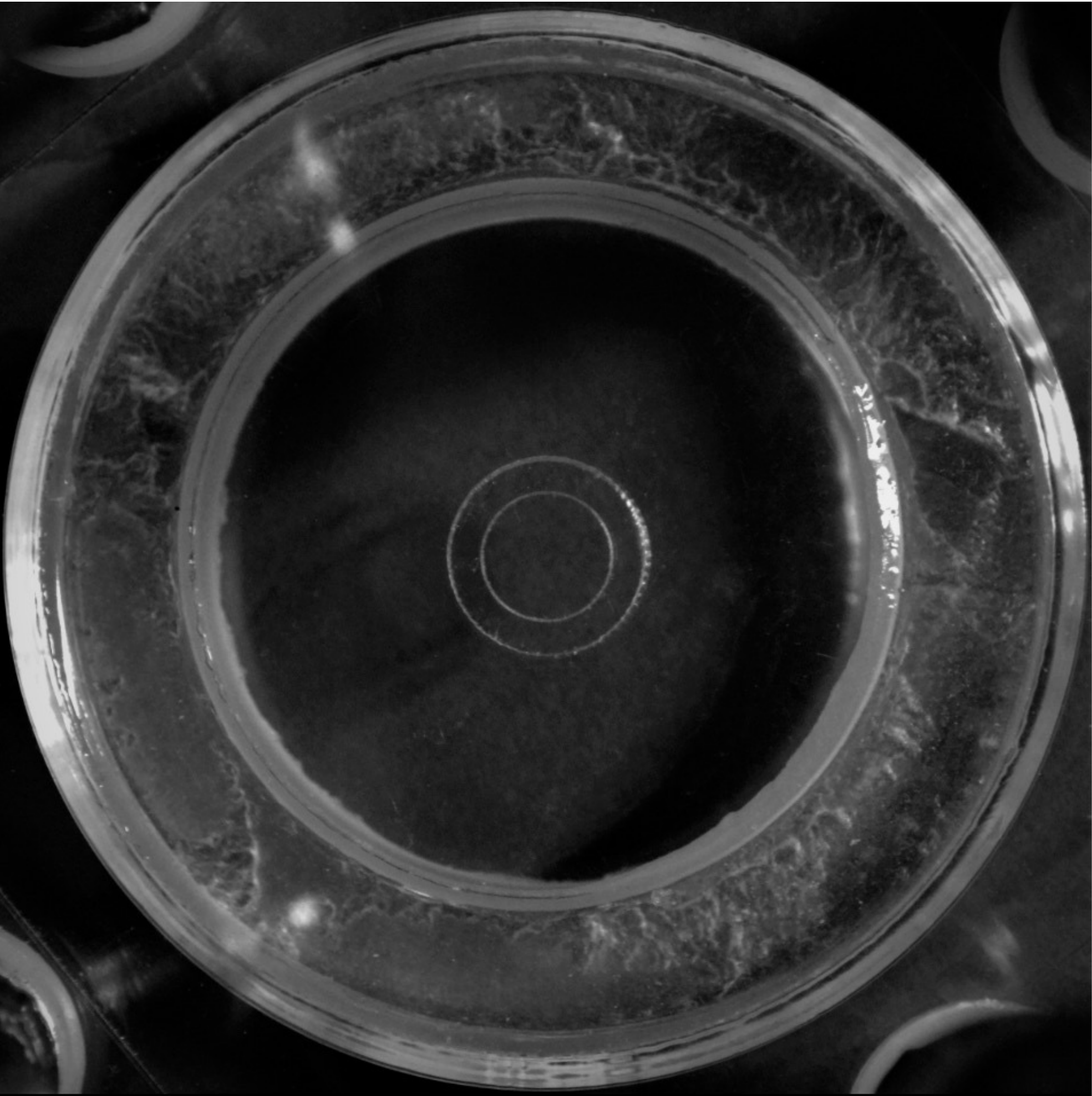}{\label{fig9a}}}
	\subfigure[Initial wavy pattern.]
         {\includegraphics[scale=0.2]{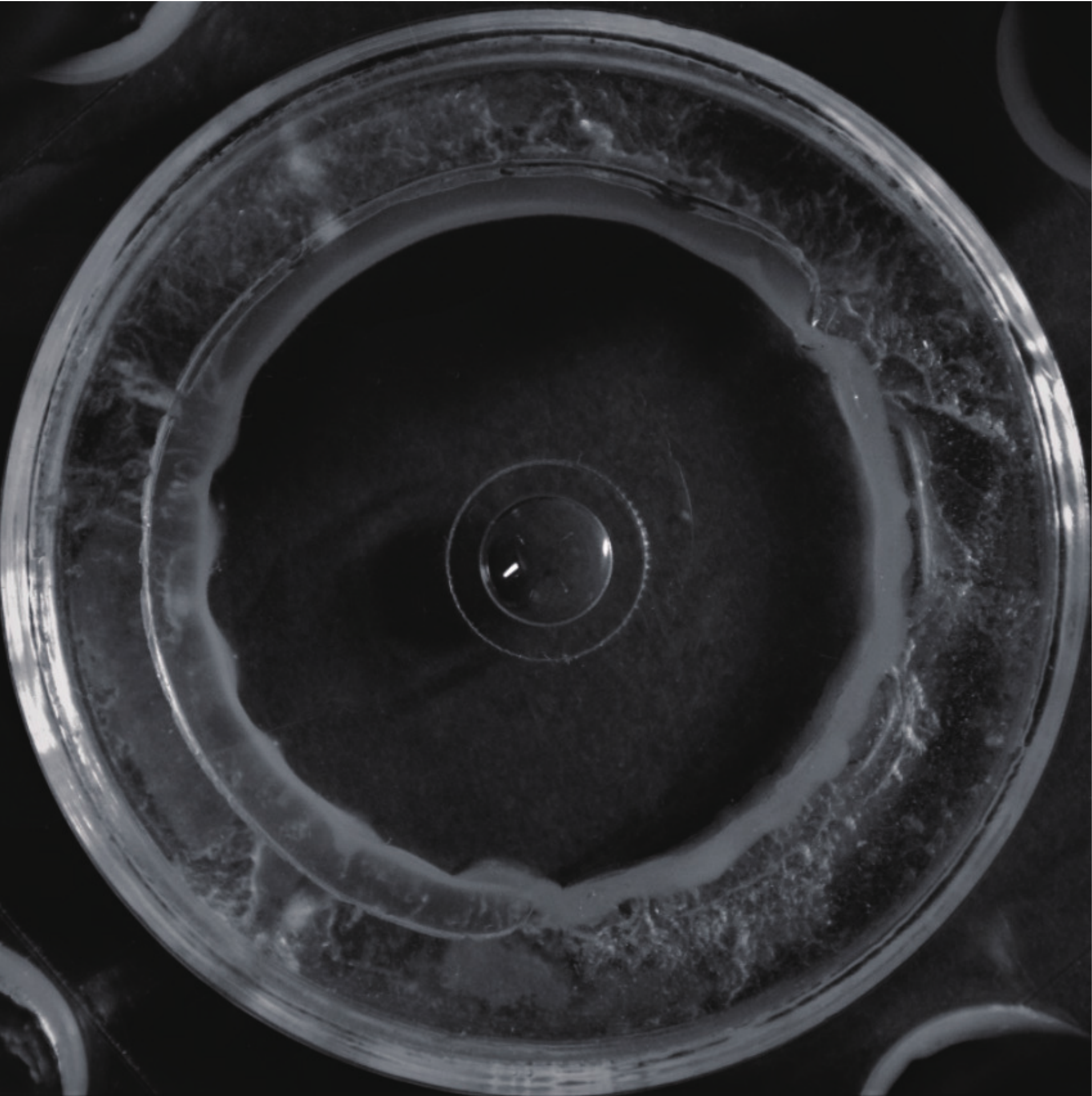}{\label{fig9b}}}
	\subfigure[Deep wrinkles.]
	{\includegraphics[scale=0.2]{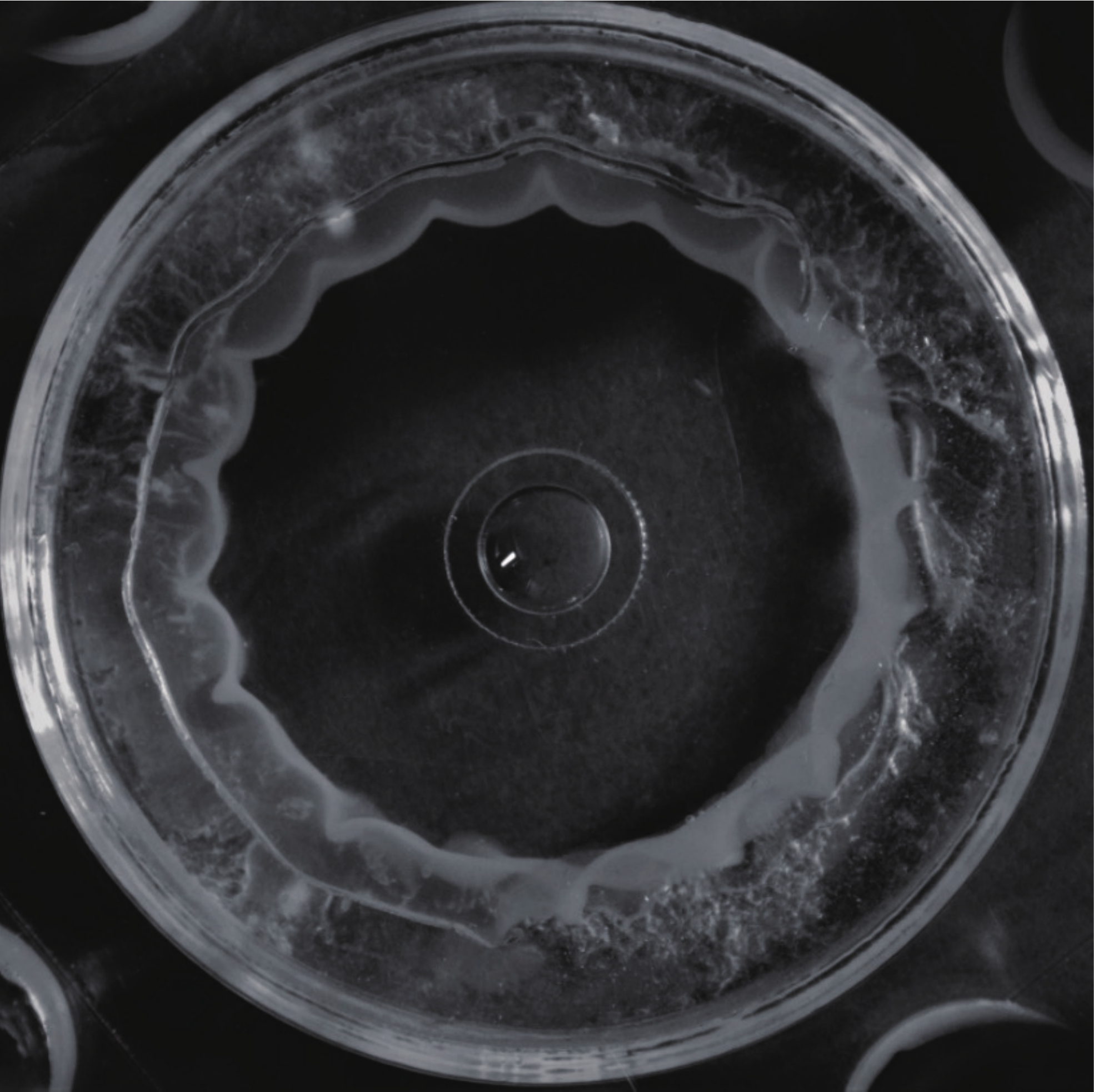}{\label{fig9c}}}
	\subfigure[Creasing mode.]
	{\includegraphics[scale=0.2]{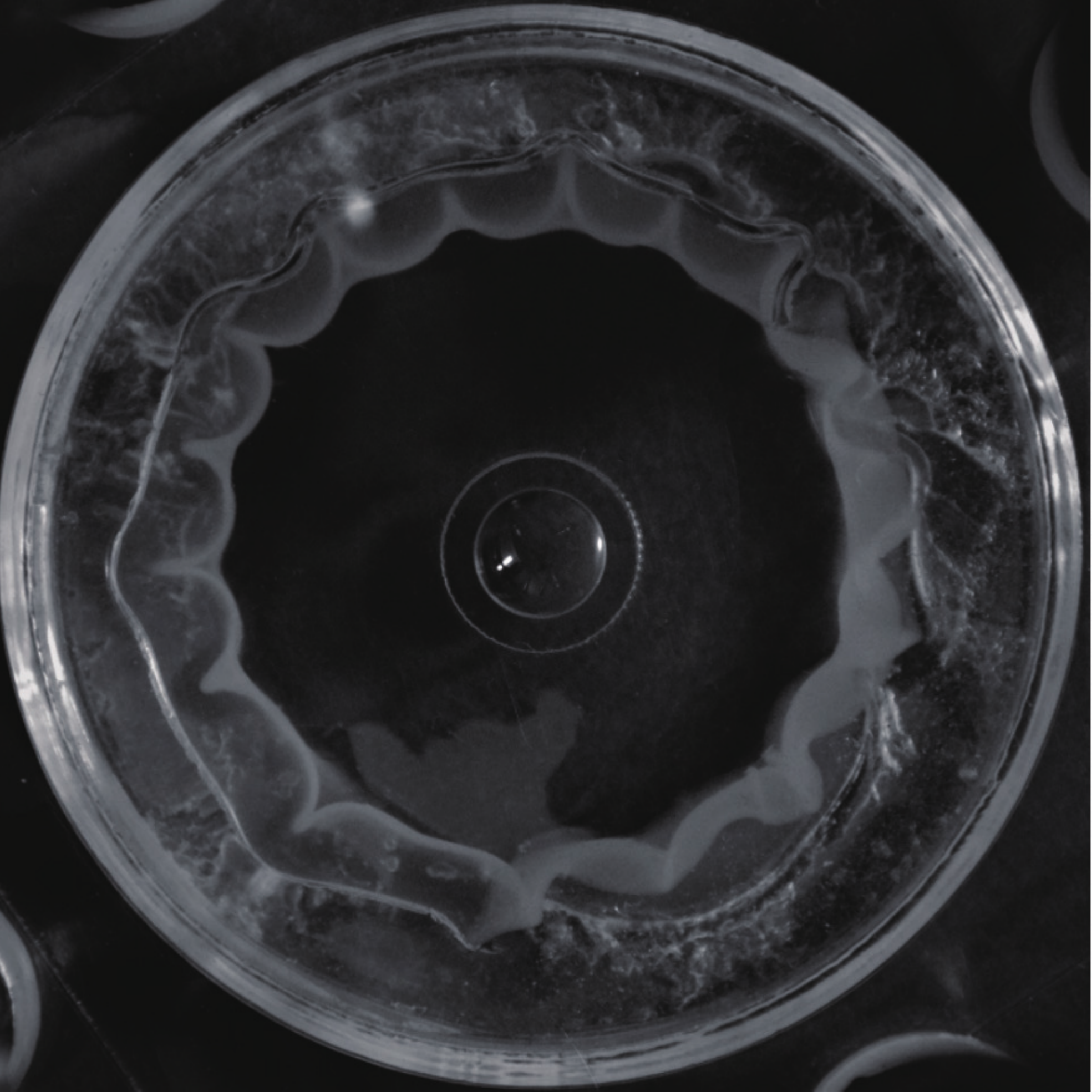}{\label{fig9d}}}
	\caption{Swelling-induced surface instabilities for sample $sp_4$. The dimensionless parameters are identified from Table \ref{table2} by $A^*=0.69$, $B^*=0.7$, and $\beta\approx6.35$. The first bifurcation produced a wavy pattern where the wavenumber was 18.}
	\label{fig9}
\end{figure}

Subsequently, the modulus ratio $\beta$ is varied, and we summarize the experimental results for samples $sp_3$ and $sp_4$ in Figures \ref{fig8} and \ref{fig9}, respectively. The modulus ratio was specified by $\beta\approx6.35$. The exact geometrical parameters for these two specimens can be found in Table \ref{table2}. It can be seen that a sinusoidal profile with 11 wrinkles were generated by swelling in \ref{fig8b}. This implies that a stiffer inner layer will reduce the wavenumber. However, differing in the unusual period-doubling in Figure \ref{fig6c}, it is found from Figures \ref{fig8c} and \ref{fig8d} that the sinusoidal morphology eventually developed into a creasing mode with surface self-contact. In addition, the results illustrated in Figure \ref{fig9} for sample $sp_4$ are comparable to the counterparts of $sp_3$. A primary deformation can be seen first while a wavy pattern was set off following progressive deformation, as presented in Figures \ref{fig9a} and \ref{fig9b}, respectively. Furthermore, the wavenumber in $sp_4$ was nearly 18. Hence the conclusion that a thinner inner layer generates a greater number of wrinkles is consistent with existing theoretical predictions \cite{25,33}. As mentioned earlier, $\beta>5.8$ renders a necessary condition for the appearance of a normal period-doubling secondary. In this case, the modulus ratio was $\beta\approx6.35$, which is marginally higher than this critical value. However, it is further found that the period-doubling secondary bifurcation needs an extremely large strain when $\beta$ is slightly greater than 5.8 such that it may give way to other modes \cite{61}. Our experimental result indicates that the creasing mode may occur as a result of the evolution of surface wrinkles when $\beta\approx6.35$, and the number of crease is identical to that of wrinkles. 

\begin{figure}[!h]
	\centering
	\subfigure[Primary deformation.]
	{\includegraphics[scale=0.3]{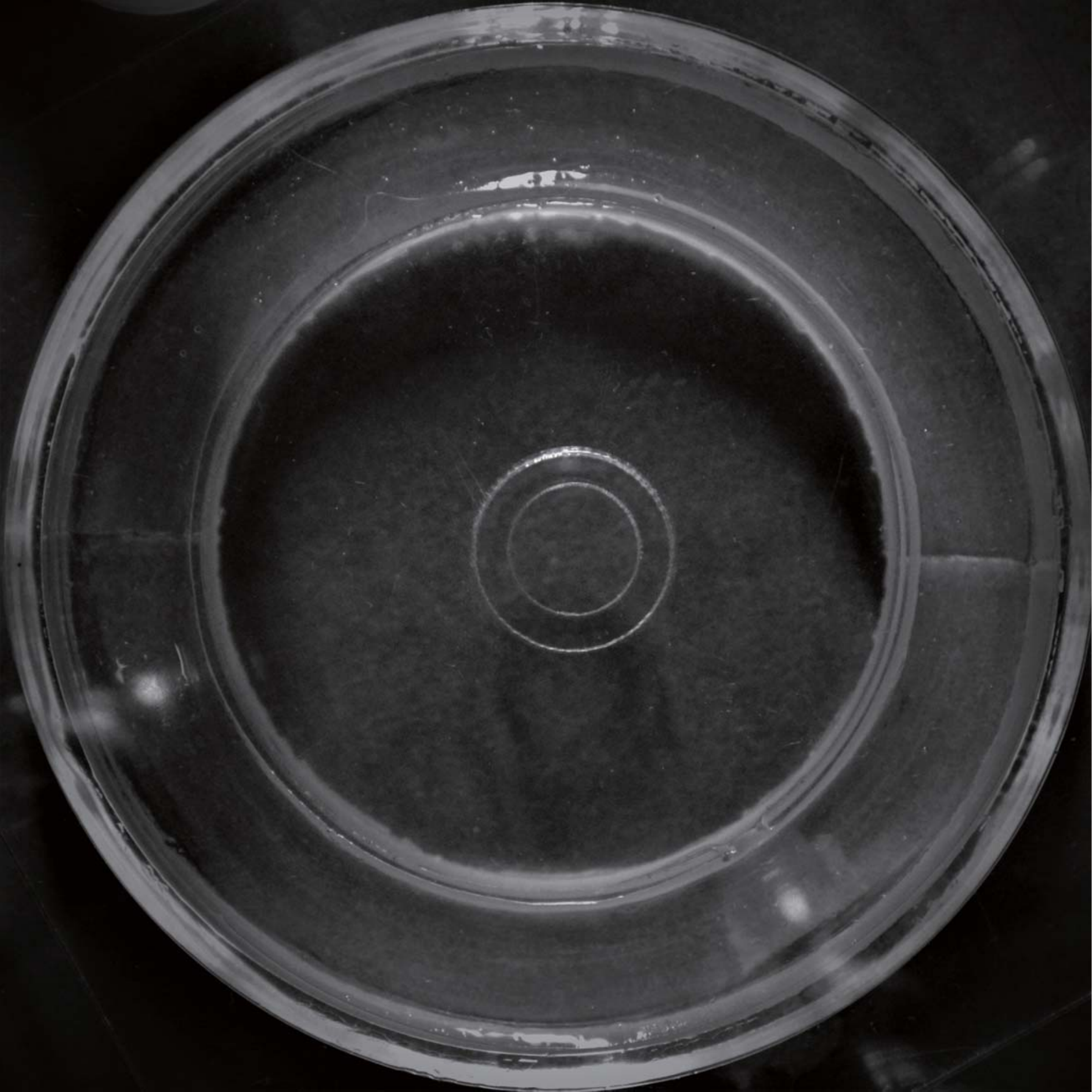}{\label{fig10a}}}
	\hspace{5mm}
	\subfigure[First bifurcation produced creasing instability.]
         {\includegraphics[scale=0.3]{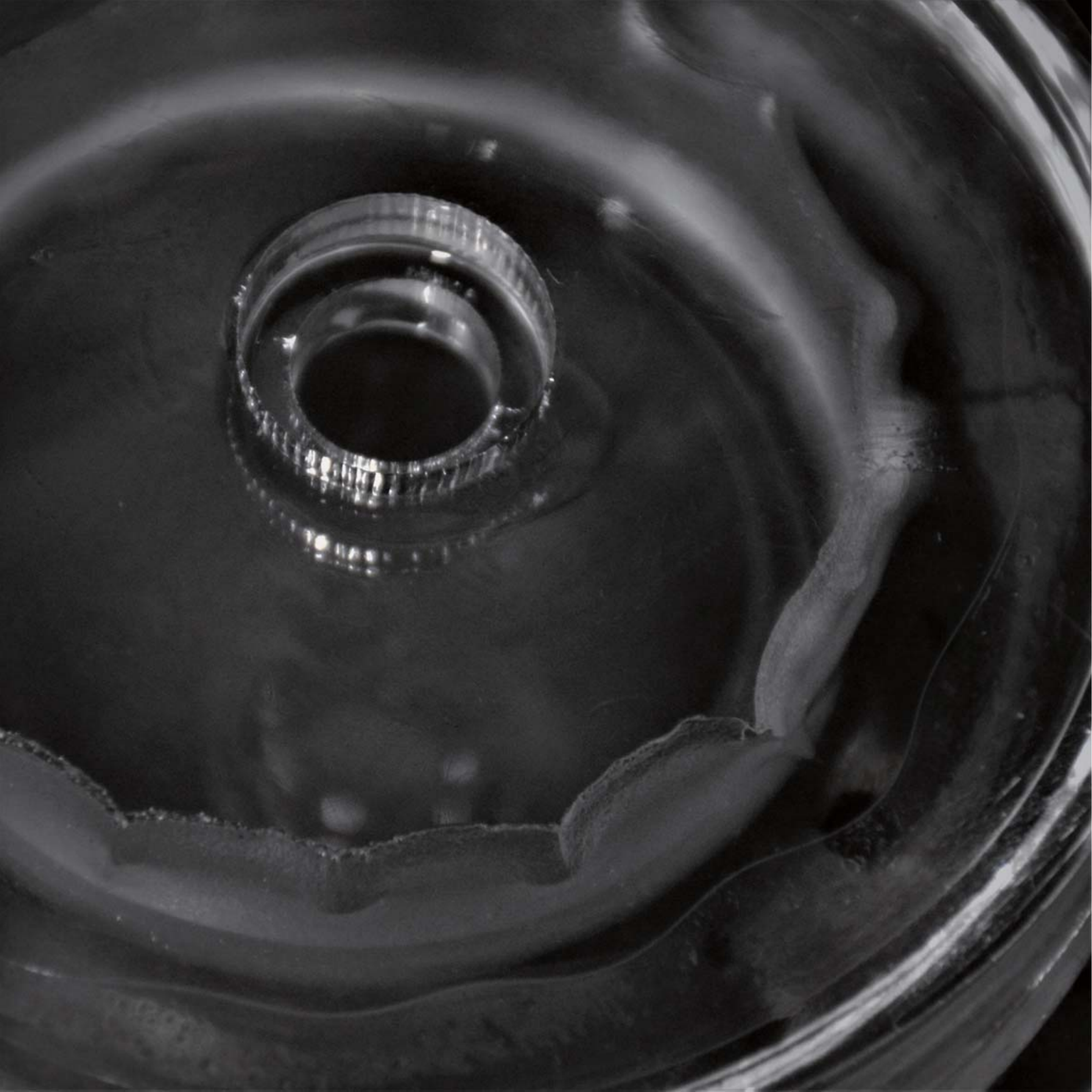}{\label{fig10b}}}
	\caption{Swelling-induced surface instabilities for sample $sp_5$. The dimensionless parameters are identified from Table \ref{table1} by $A^*=0.67$, $B^*=0.7$, and $\beta\approx1$. }
	\label{fig10}
\end{figure}

Finally, we depict the experimental consequence for $sp_5$ in Figure \ref{fig10}, where the geometrical size was consistent with that of $sp_1$ and $sp_3$. Yet the modulus ratio was practically equal to unity. In a previous study by Cai and Fu \cite{57}, a weakly nonlinear analysis for planar film-substrate structures, giving rise to the amplitude equation of wrinkling mode, was performed, and it turns out that the bifurcation is subcritical if $\beta<\beta_c=1.74$. Afterwards, Jin el al. \cite{34} found that the critical value of $\beta$ where the bifurcation nature transforms is dependent on the geometrical parameters and ranges from $1.18$ to $1.6$. For a subcritical bifurcation, it is usually expected that creasing mode may happen instead of wrinkling mode. Therefore, the last specimen is used to unveil the actual surface pattern for a subcritical bifurcation. Seen from Figure \ref{fig10}, swelling primarily induced an axisymmetric deformation when $\beta\approx1$. However, the surface of the inner layer was self-contact as the hoop stress exceeded a critical value, forming a creasing morphology, as displayed in detail in Figure \ref{fig10b}. To clearly depict the creases, a sideway view is shown in Figure \ref{fig10b}. In addition, no other patterns were perceived in our experiment. 

In this section, we have carried out an experimental investigation using five fabricated samples. It is found that the primary bifurcation may create a wavy pattern concentrated on the inner layer unless the modulus ratio $\beta\approx 1$ where a creasing mode occurs instead. Meanwhile, an interfacial wrinkling was discovered. Although this special instability is out of the scope of this study, it still can motivate a further study on the competition between surface wrinkles and interfacial ones. Furthermore, a winkle to crease transition and a wrinkle to period-doubling transition were seen. In the following sections, the experimental results will be employed to validate the theoretical model of volumetric growth by comparisons among experimental findings, theoretical predictions, and finite element simulations. 

\section{Modelling}
Although the incompressible neo-Hookean model is applied in the swelling experiments, we decide to establish a theoretical model for constrained growth and derive the bifurcation condition using volumetric growth theory without specifying an exact form of the strain energy function.
\subsection*{Growth theory and basic equations}
\begin{figure}[!h]
    \centering\includegraphics[width=5.5in]{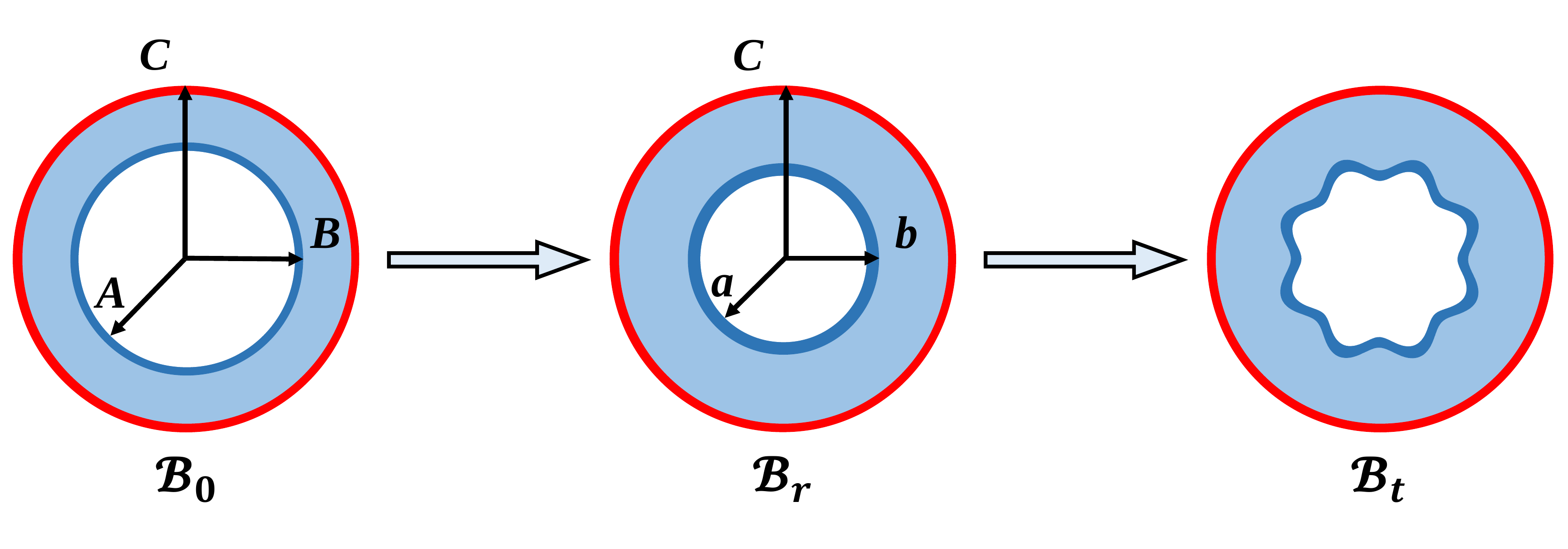}
    \caption{(Color online) Sketches of the initial state, the basic state and the wrinkled state of a bilayered tubular tissue where the outer boundary is restricted subjected to growth.}
    \label{fig11}
\end{figure}

In our experiments, both layers of the bilayered tubular structure are modeled by incompressible hyperplastic materials. Meanwhile, the outer boundary is assumed to be fixed in the growth process while the interface keeps perfectly bonded. As illustrated in Figure \ref{fig11}, the initial inner, interfacial, and outer radii are denoted by $A$, $B$, and $C$, respectively, in the reference configuration $\mathcal{B}_0$. Note that this is in accordance with the notations used in our experiment. Under growth, the tubular tissue will grow thicker, where the inner and interfacial radii become $a$ and $b$ in the current configuration $\mathcal{B}_r$. As the growth factor reaches a critical value, surface wrinkling will emerge in the inner layer and the bifurcated state is called $\mathcal{B}_t$. We add a hat on a quantity if it belongs to the outer layer or otherwise it is owned by the inner layer. For example, the strain energy function for the inner layer is represented by $W$ while the one for the outer layer both layers is written as $\hat{W}$. All derivations of the governing equations are similar for both layers, hence only the procedure for the inner layer will be shown for brevity. 

For convenience, the cylindrical polar coordinates system is adopted in both the reference and the current configurations and the common orthonormal basis reads $\{{\bm e}_r, {\bm e}_\theta, {\bm e}_z \}$. The coordinates of a representative material point in $\mathcal{B}_0$ and $\mathcal{B}_r$ are traditionally described by $(R,\Theta,Z)$ and $(r,\theta,z)$, respectively. Considering that the primary deformation from $\mathcal{B}_0$ to $\mathcal{B}_r$ is axisymmetric, the deformation gradient for the inner layer is given by
\begin{equation}
    \mathbf{F}=\lambda_r\bm e_r\otimes\bm e_r+\lambda_\theta\bm e_\theta\otimes\bm e_\theta+\lambda_z\bm e_z\otimes\bm e_z,
    \label{eq5}
\end{equation}
where $\lambda_r$, $\lambda_\theta$, and $\lambda_z$ stand for the principal stretches in the corresponding directions. In particular, we have $\lambda_r=\mathrm{d} r/\mathrm{d} R$ and $\lambda_\theta={r}/{R}$, and the principal stretch in the $z$-direction is reduced to unity for a plane-strain deformation such that  $\lambda_z=1$. According to the theory of volumetric growth \cite{17}, the deformation gradient can be decomposed into
\begin{equation}
    \mathbf{F}=\mathbf{A}\mathbf{G},
    \label{eq6}
\end{equation}
where $\mathbf{A}$ is an elastic deformation tensor and $\mathbf{G}$ is a growth tensor describing the addition or diminution of materials. 

Assuming that the growth tensor is diagonal, the following expressions can be found:
\begin{equation}
    \mathbf{G}=g_r\bm e_r\otimes\bm e_r+g_\theta\bm e_\theta\otimes\bm e_\theta+\bm e_z\otimes\bm e_z,,~~\mathbf{A}=\alpha_r\bm e_r\otimes\bm e_r+\alpha_\theta\bm e_\theta\otimes\bm e_\theta+\bm e_z\otimes\bm e_z,
    \label{eq7}
\end{equation}
where $g_i$ ($i=1,2$) is the growth factors, and $\alpha_i$ ($i=1,2$) denotes the radius-dependent elastic principal stretch. Here and hereafter, the index $i=1$ corresponds to the $r$-direction while $i=2$ corresponds to the $\theta$-direction, respectively. If $g_i=1$ there is no change in volume in the $i$th direction, and $g_i>1$ or $g_i<1$ means that there is a growth or an atrophy in the $i$th direction. Further, we could write $\alpha_r=g_r^{-1}\mathrm{d} r/\mathrm{d} R$ and $\alpha_\theta=g_\theta^{-1}{r}/{R}$. In view of the elastic incompressibility $\operatorname{det}\mathbf{A}=1$, the boundary condition at $r=C$ and the displacement continuity condition at $r=b$, we could characterize the deformation in the basic state $\mathcal{B}_r$ by
\begin{equation}
\begin{aligned}
    &r =\sqrt{B^2(\hat{g}_r\hat{g}_\theta-g_r g_\theta)+C^2 \left(1-\hat{g}_r\hat{g}_\theta \right)+g_rg_\theta R^2},~~~a< r<b,\\
    &r=\sqrt{C^2 \left(1-\hat{g}_r\hat{g}_\theta\right)+\hat{g}_r\hat{g}_{\theta }R^2},\hspace{32mm}b<r<C.
    \label{eq8}
\end{aligned}
\end{equation}

In terms of the strain energy function $W(\mathbf{A})$, or equivalently in terms of the principal stretches $W(\alpha_r,\alpha_\theta)$, the Cauchy stress tensor $\bm{\sigma}$ is given by \cite{16}
\begin{equation}
    \bm{\sigma}=\mathbf{A}\frac{\partial W}{\partial \mathbf{A}}-p\mathbf{I},
    \label{eq9}
\end{equation}
where $p$ is the hydrostatic pressure (see also equation (\ref{eq2})) and $\mathbf{I}$ the second-order identity tensor.

In the absence of the body force, we arrive at the equilibrium equation
\begin{equation}
    \operatorname{div} \bm \sigma=\mathbf{0},
    \label{eq10}
\end{equation}
where ``$\mathrm{div}$'' stands for the divergence operator evaluated in the current configuration. In component form, the only equation that is not automatically satisfied writes
\begin{equation}
    \dfrac{\mathrm{d}\sigma_{rr}}{\mathrm{d} r}+\dfrac{\sigma_{rr}-\sigma_{\theta\theta}}{r}=0.
    \label{eq11}
\end{equation}
The traction-free condition at the inner surface and the continuity condition yield
\begin{equation}
\begin{aligned}
    &\sigma_{rr}=0,\hspace{5.5mm}\mathrm{on}~~r=a,\\
    &\sigma_{rr}=\hat{\sigma}_{rr},~~~\mathrm{on}~~r=b.
    \label{eq12}
\end{aligned}
\end{equation}

Next, we make use of (\ref{eq9}), (\ref{eq11}) and apply the boundary condition and continuity condition (\ref{eq12}) to obtain the expressions of the stress and hydrostatic pressures as follows
\begin{equation}
\begin{aligned}
    &\sigma_{rr}=\int_a^r\frac{1}{r}\left(\alpha_{r} W_{, 1}-\alpha_{\theta} W_{, 2}\right) \mathrm{d} r,\\
    &p=\alpha_{r} W_{, 1}+\int_{a}^{r} \frac{1}{r}\left(\alpha_{r} W_{, 1}-\alpha_{\theta} W_{, 2}\right) \mathrm{d} r,\\
    &\hat{p}=\hat{\alpha}_{r} \hat{W}_{, 1}+\int_{b}^{r} \frac{1}{r}\left(\hat{\alpha}_{r} \hat{W}_{, 1}-\hat{\alpha}_{\theta} \hat{W}_{, 2}\right) \mathrm{d} r +\int_{a}^{b} \frac{1}{r}\left(\alpha_{r} W_{, 1}-\alpha_{\theta} W_{, 2}\right) \mathrm{d} r,
    \label{eq13}
\end{aligned}
\end{equation}
where a comma behind a quantity indicates differentiation with respect to the corresponding variable, e.g. $W_{, 1}=\partial W / \partial \alpha_{r}$. 

\subsection*{Linearized incremental equation}
It is appropriate to formulate the linearized incremental equation for further bifurcation analysis following the procedure in \cite{16,76}. To this end, we put a tilde on a symbol to depict that it is evaluated in $\mathcal{B}_t$, for instance, the position vector in $\mathcal{B}_t$ is denoted by $\widetilde{\bm{x}}$. Again, we only supply derivations for the inner layer as well. The position $\widetilde{\bm{x}}$ is attained by superimposing an infinitesimal displacement field $\delta \bm{x}$ on $\mathcal{B}_r$, and this field is given by
\begin{equation}
    \delta \bm x=u(r,\theta)\bm e_r+v(r,\theta)\bm e_\theta,
    \label{eq14}
\end{equation}
where $u(r,\theta)$ and $v(r,\theta)$ are the incremental displacements in the radial and hoop directions. Then the deformation gradient arising from $\mathcal{B}_0\longrightarrow\mathcal{B}_t$ can be expressed as $\widetilde{\mathbf{F}}=(\mathbf{I}+\bm{\eta})\mathbf{F}$, with $\bm{\eta}$ given by
\begin{align}
    &\bm\eta=\dfrac{\partial u}{\partial r}\bm e_r\otimes\bm e_r+\dfrac{1}{r}\left(\dfrac{\partial u}{\partial \theta}-v\right)\bm e_r\otimes\bm e_\theta+\dfrac{\partial v}{\partial r}\bm e_\theta\otimes\bm e_r+\dfrac{1}{r}\left(\dfrac{\partial v}{\partial \theta}+u\right)\bm e_\theta\otimes\bm e_\theta.
    \label{eq15}
\end{align}

The linearized incompressibility condition requires
\begin{equation}
    \operatorname{tr}{\bm\eta}=\dfrac{\partial u}{\partial r}+\dfrac{1}{r}\left(\dfrac{\partial v}{\partial \theta}+u\right)=0,
    \label{eq16}
\end{equation}
where ``$\mathrm{tr}$'' is the trace operator. 

To construct the incremental stress, we denote the nominal stress in $\mathcal{B}_r$ by $\mathbf{S}$ and the counterpart in $\mathcal{B}_r$ by $\widetilde{\mathbf{S}}$. In particular, the $\mathbf{S}$ can be deduced from the identity $\mathbf{S} =J\mathbf{F}^{-1}\bm \sigma$ with $J=\operatorname{det} \mathbf{F}=\operatorname{det} \mathbf{G}$ depicting the volume change. Referring to \cite{35,43}, we define the following incremental stress tensor by use of the nominal stresses:
\begin{equation}
    \bm\chi^\mathrm{T}={J}^{-1}\mathbf{F}(\widetilde{\mathbf{S}}-\mathbf{S}),
    \label{eq17}
\end{equation}
where the superscript ``$\mathrm{T}$'' represents transpose.  Next, taking the Taylor expansion of $\bm\chi$ in $\mathbf{F}$ and keeping all linear terms furnish
\begin{equation}
    \chi_{ij}=\mathcal{A}_{jilk}\eta_{kl}+p\eta_{ji}-p^{*}\delta_{ji},
    \label{eq18}
\end{equation}
where $p$ has been given in (\ref{eq13}), $p^{*}$ is the corresponding incremental counterpart, and $\mathcal{A}_{jilk}$ is the first-order instantaneous modulus and takes the following formula \cite{35,68,76}:
\begin{align*}
    &\mathcal{A}_{iijj}=\mathcal{A}_{jjii}=\alpha_i\alpha_jW_{,ij},~~\mathrm{no~summation~on~}i~\mathrm{or}~j,\notag\\
    &\mathcal{A}_{ijij}=\dfrac{\alpha_iW_{,i}-\alpha_jW_{,j}}{\alpha_i^2-\alpha_j^2}\alpha_i^2,~~\alpha_i\neq \alpha_j,~~\mathrm{no~summation~on~}i~\mathrm{or}~j,\notag\\
    &\mathcal{A}_{ijji}=\mathcal{A}_{ijij}-\alpha_iW_{,i},~~i\neq j,~~\mathrm{no~summation~on~}i~\mathrm{or}~j.
\end{align*}

The incremental equilibrium equation for the inner layer can be written as
\begin{equation}
    \operatorname{div}\bm\chi^\mathrm{T}=\bm 0,
    \label{eq19}
\end{equation}
or in component form:
\begin{equation}
\begin{aligned}
    &\dfrac{\partial \chi_{rr}}{\partial r}+\dfrac{1}{r}\dfrac{\partial \chi_{r \theta}}{\partial \theta}+\dfrac{\chi_{rr}-\chi_{\theta \theta}}{r}=0,\\
    &\dfrac{\partial  \chi_{\theta r}}{\partial r}+\dfrac{1}{r}\dfrac{\partial\chi_{\theta\theta}}{\partial \theta}+\dfrac{\chi_{\theta r}+\chi_{r\theta}}{r}=0.
    \label{eq20}
\end{aligned}
\end{equation}
Furthermore, the incremental boundary conditions and continuity conditions can be expressed by
\begin{equation}
\begin{aligned}
    &\bm{\chi} \bm{e}_r \big|_{r=a}=\bm 0,\\
    &(\bm{\chi}-{\hat{\bm\chi}}) \bm{e}_r \big|_{r=b}=\bm 0,~~~(u-\hat{u}) \big|_{r=b}=(v-\hat{v}) \big|_{r=b}=0,\\
    &\hat{u} \big|_{r=C}=\hat{v} \big|_{r=C}=0.
    \label{eq21}
\end{aligned}
\end{equation}

Currently, the incremental equation as well as the boundary conditions and continuity conditions for further bifurcation analysis are established. In particular, equation (\ref{eq20}) contains spacial-dependent coefficients. In many previous analysis, such an eigenvalue problem arising from (\ref{eq16}) and (\ref{eq20}) associated with (\ref{eq21}) has been solved using determinant method \cite{24,33}. Notwithstanding, in the next section, the Stroh method \cite{46,47,48,49,50,77,78} will be utilized to derive the bifurcation condition in a more compact way.

\section{Stroh formulation and the surface impedance matrix method}
Here we use Stroh formulation and the surface impedance matrix method to solve the eigenvalue problem of the linearized incremental system. For bilayer or multilayer models, the surface impedance matrix could express the bifurcation condition in a succinct way, and it is convenient to deal with continuity conditions on the interface. It should be pointed out that the surface impedance matrix method has been well formulated for traction boundary conditions in literature. Yet for the current problem where a displacement boundary condition is involved, we shall slightly modify the classical method and then apply it to carry out a bifurcation analysis. Moreover, we just present the main procedures for the inner layer while the counterparts for the outer layer can be derived in a similar way and will be directly written down when necessary.

We seek the solution of equation (\ref{eq20}) in the following form
\begin{equation}
    u(r,\theta)=U(r) \textrm{cos}(n\theta),~~~v(r,\theta)=V(r)\textrm{sin}(n\theta),~~~p^{*}(r,\theta)=P(r)\textrm{cos}(n\theta),
   \label{eq22}
\end{equation}
where $n$ is called the circumferential wavenumber, $U$, $V$, and $P$ are unknown functions of $r$. Similarly, we can express the components of the incremental stress tensor by
\begin{align}
    \chi_{rr}(r,\theta)=X_{rr}(r) \mathrm{cos}(n\theta),~~~\chi_{r\theta}(r,\theta)=X_{r\theta}(r)\mathrm{sin}(n\theta),
    \label{eq23}
\end{align}
where $X_{ij}$ are functions to be determined. It is shown later that these formulations will greatly simplify the bifurcation analysis.

In light of the incompressibility condition (\ref{eq16}), we can obtain the relation between $U(r)$ and $V(r)$ as follows
\begin{equation}
    n V(r)+r \frac{\mathrm{d}U(r)}{\mathrm{d} r}+U(r)=0.
    \label{eq24}
\end{equation}
Meanwhile, solving $P(r)$ from the expression of $\chi_{rr}$ yields
\begin{equation}
    P(r)=\frac{1}{r}\Big((\mathcal{A}_{rr\theta \theta}-\mathcal{A}_{rrrr}-p(r)) (n V(r)+U(r))-r X_{rr}(r)\Big).
    \label{eq25}
\end{equation}

Next, we define two vectors by $\boldsymbol{U}(r)=[U(r), V(r)]^\mathrm{T}$ and $ \bm X(r)=[r X_{rr}(r),r X_{r\theta}(r)]^\mathrm{T}$ and introduce the displacement-traction vector $\bm \xi(r)$ given by
\begin{align}
    \bm \xi(r)=\left[\bm U(r),\bm X(r)\right]^\mathrm{T}.
    \label{eq26}
\end{align}
It is then possible to derive a first-order differential system for $\bm \xi(r)$ by means of (\ref{eq18}), (\ref{eq20}), (\ref{eq24}) and (\ref{eq25}):
\begin{equation}
    \frac{\mathrm{d} \bm \xi(r)}{\mathrm{d} r}=\frac{1}{r} \mathbf{Q}(r) \bm \xi(r),~~a<r<b,
    \label{eq27}
\end{equation}
which is referred to as the $Stroh$ $formulation$ of the incremental problem \cite{77}. In the above equation, $\mathbf{Q}(r)$ is the so-called Stroh matrix which admits the following block representation
\begin{equation}
    \mathbf{Q}(r)=\left[\begin{array}{cc}
    \mathbf{Q}_{1} & \mathbf{Q}_{2} \\
    \mathbf{Q}_{3} & \mathbf{Q}_{4}
    \end{array}\right],
    \label{eq28}
\end{equation}
where the $2 \times 2$ sub-blocks $\mathbf{Q}_2$ and $\mathbf{Q}_3$ are real and symmetric, and $\mathbf{Q}_4=-\mathbf{Q}_1^\mathrm{T}$. In particular, the matrices $\mathbf{Q}_1$ and $\mathbf{Q}_2$ read
\begin{align}
	&\mathbf{Q}_1(r)=\left[\begin{array}{cc}
	-1 & -n \\
	\dfrac{n \left(\mathcal{A}_{r \theta \theta r}+p\right)}{\mathcal{A}_{r \theta r \theta }} & \dfrac{\mathcal{A}_{r \theta \theta r}+p}{\mathcal{A}_{r \theta r \theta }}
	\end{array}\right], \quad
	\mathbf{Q}_2(r)=\left[\begin{array}{cc}
	0 & 0 \\
	0 & \dfrac{1}{\mathcal{A}_{r \theta r \theta }}
	\end{array}\right], \label{eq29}
\end{align}
and the non-zero components of $\mathbf{Q}_3$ are given by
\begin{equation}
\begin{aligned}
    &(\mathbf{Q}_{3})_{11}=\mathcal{A}_{\theta \theta \theta \theta }-\mathcal{A}_{\theta \theta rr}+n^2 \left(\mathcal{A}_{\theta r\theta r}-\frac{\left(\mathcal{A}_{\theta r r\theta }+p\right) \left(\mathcal{A}_{r \theta \theta r }+p\right)}{\mathcal{A}_{r \theta r \theta }}\right)-\mathcal{A}_{rr\theta \theta}+\mathcal{A}_{rrrr}+2 p,\\
    &(\mathbf{Q}_{3})_{12}=(\mathbf{Q}_{3})_{21}=n \left(\mathcal{A}_{\theta \theta \theta \theta }-\mathcal{A}_{\theta \theta rr}+\mathcal{A}_{\theta r\theta r}-\frac{\left(\mathcal{A}_{\theta rr\theta }+p\right) \left(\mathcal{A}_{r\theta \theta r}+p\right)}{\mathcal{A}_{r\theta r\theta }}-\mathcal{A}_{rr\theta \theta }+\mathcal{A}_{rrrr}+2 p\right),\\
    &(\mathbf{Q}_{3})_{22}=\mathcal{A}_{\theta r\theta r}+n^2 \left(\mathcal{A}_{\theta \theta \theta \theta }-\mathcal{A}_{\theta \theta rr}-\mathcal{A}_{rr\theta \theta }+\mathcal{A}_{rrrr}+2 p\right)-\frac{\left(\mathcal{A}_{\theta rr\theta }+p\right) \left(\mathcal{A}_{r\theta \theta r}+p\right)}{\mathcal{A}_{r\theta r\theta }}.
    \label{eq30}
\end{aligned}
\end{equation}

We resort to the $impedance~matrix~method$ \cite{50,77,78} to solve the incremental elastic problem and define the $4\times4$ matrix $\mathbf{M}(r, r_i)$ that satisfies $\mathbf{M}\left(r_{i}, r_{i}\right)=\mathbf{I}_{(4)}$ where $\mathbf{I}_{(4)}$ denotes the $4\times4$ identity matrix as the solution of the initial value problem 
\begin{align}
\frac{\mathrm{d} \mathbf{M}\left(r, r_{i}\right)}{\mathrm{d} r}=\frac{1}{r} \mathbf{Q}(r) \mathbf{M}\left(r, r_{i}\right),~~a<r<b,
    \label{eq31}
\end{align}
where $r_i$ is a constant lying in $(a,b)$. We further rewrite $\mathbf{M}(r, r_i)$ in a block representation 
\begin{align}
\mathbf{M}\left(r, r_{i}\right)=\left[\begin{array}{cc}
		\mathbf{M}_{1}(r,r_{i}) & \mathbf{M}_{2}(r,r_{i}) \\
		\mathbf{M}_{3}(r,r_{i}) & \mathbf{M}_{4}(r,r_{i})
		\end{array}\right].
		\label{eq32}
\end{align}

Subsequently, we suppose that the traction vector $\bm X(r)$ and the displacement vector $\bm U(r)$ are connected by
\begin{equation}
    \bm X(r)=\mathbf{Z}(r) \bm U(r).
    \label{eq33}
\end{equation}
In the above formula, the $\mathbf{Z}(r)$ corresponds to a surface impedance matrix. Bearing in mind that each column of  $\mathbf{M}(r, r_i)$ is a solution of equation (\ref{eq27}), it can be deduced from the traction-free boundary condition on $r=a$ that $\mathbf{Z}=\mathbf{M}_3 \mathbf{M}_1^{-1}$. In addition, substituting (\ref{eq26}) and (\ref{eq33}) into equation (27) and eliminating the dependence of $\bm U$ results in a \textit{Riccati equation} for $\mathbf{Z}$:
\begin{equation}
    \frac{d \mathbf{Z}(r)}{d r}=\frac{1}{r}\left(\mathbf{Q}_{3}-\mathbf{Z} \mathbf{Q}_{1}-\mathbf{Z} \mathbf{Q}_{2} \mathbf{Z}+\mathbf{Q}_{4} \mathbf{Z}\right),~~a<r<b.
    \label{eq34}
\end{equation}


This matrix equation is subjected to the boundary condition at $r=a$, which implies $\mathbf{Z}(a)=\bm 0$, and the continuity condition at $r=b$. Note that a displacement boundary condition is imposed at $r=C$. To derive a bifurcation condition,   instead of (\ref{eq33}) we assume 
\begin{equation}
    \hat {\bm U}(r)=\mathbf{K}(r) \hat {\bm X}(r).
    \label{eq35}
\end{equation}
Referring to the fixed boundary condition $\hat{\bm U}(C)=\bm 0$, it is found that $\mathbf{K}=\hat{\mathbf{M}}_2 \hat{\mathbf{M}}_4^{-1}$. Furthermore, applying the similar deduction as above, we eventually acquire another \textit{Riccati equation}:
\begin{equation}
    \frac{d \mathbf{K}(r)}{d r}=\frac{1}{r}\left(\hat{\mathbf{Q}}_{2}-\mathbf{K} \hat{\mathbf{Q}}_{4}-\mathbf{K} \hat{\mathbf{Q}}_{3} \mathbf{K}+\hat{\mathbf{Q}}_{1} \mathbf{K}\right),~~b<r<C,
    \label{eq36}
\end{equation}
where the expressions of $\hat{\mathbf{Q}}_1$ to $\hat{\mathbf{Q}}_4$ can be attained by replacing variables in (\ref{eq29}) and (\ref{eq30}).

Correspondingly, the boundary condition $\mathbf{K}(C)=\bm 0$ and the continuity condition at $r=b$ are imposed to equation (\ref{eq36}). Then a bifurcation condition can be derived by use of the matching condition at $r=b$ for the two fundamental unknowns $\mathbf{Z}(r)$ and $\mathbf{K}(r)$ and takes the following form
\begin{equation}
    \operatorname{det}\left(\mathbf{Z}(b)\mathbf{K}(b)-\mathbf{I}\right)=0.
    \label{eq37}
\end{equation}

The bifurcation condition (\ref{eq37}) can be solved numerically in software \textit{Mathematica} \cite{79} to identify the onset of surface wrinkling and the associated surface pattern for a given material model and for specified parameters. It turns out that the impedance matrix method is more efficient compared with the determinant method. In particular, it offers a more elegant formulation for all derivations as well as the bifurcation condition, without a specific manipulation of the continuity condition in layered structures. Furthermore, the continuity conditions are satisfied at any position for a monolayer structure, so the bifurcation analysis presented earlier can be applied to other problems where a displacement boundary condition exists.

It is pointed out that, for the current problem, a detailed bifurcation analysis was performed by Li et al. \cite{24} and Moulton and Goriely \cite{27}, and an asymptotic analysis was conducted by Jin et al \cite{33}. Therefore, the influence of the geometrical and material parameters on the initiation of a wavy pattern has been clearly revealed and it is unnecessary to perform parametric analysis of the buckling condition here. However, this study aims to validate the effectiveness of the volumetric growth model in explaining pattern formation and evolution in tubular tissues and to re-derive the bifurcation condition using Stroh formulation. In the subsequent part, we shall adopt the parameters used in our experiments to make an exhaustive comparison of the sinusoidal pattern.  

We emphasize that the buckling analysis in this section allows the circumferential and radial growth factors to be varied. In our illustrative experiments, the n-Hexanes penetrate the structure through the inner surface, see the experimental setup in Figure \ref{fig1}. This in fact results in an inhomogeneous growth in the radial direction. Yet in our previous study, it has been unraveled that a growth gradient has a negligible effect on surface wrinkling and the evolution of wrinkles \cite{43}. Thus, we consider a homogeneous growth type in both layers such that $g_r=g_\theta=\hat{g}_r=\hat{g}_\theta:=g$.

\begin{figure}[!h]
    \centering\includegraphics[width=6in]{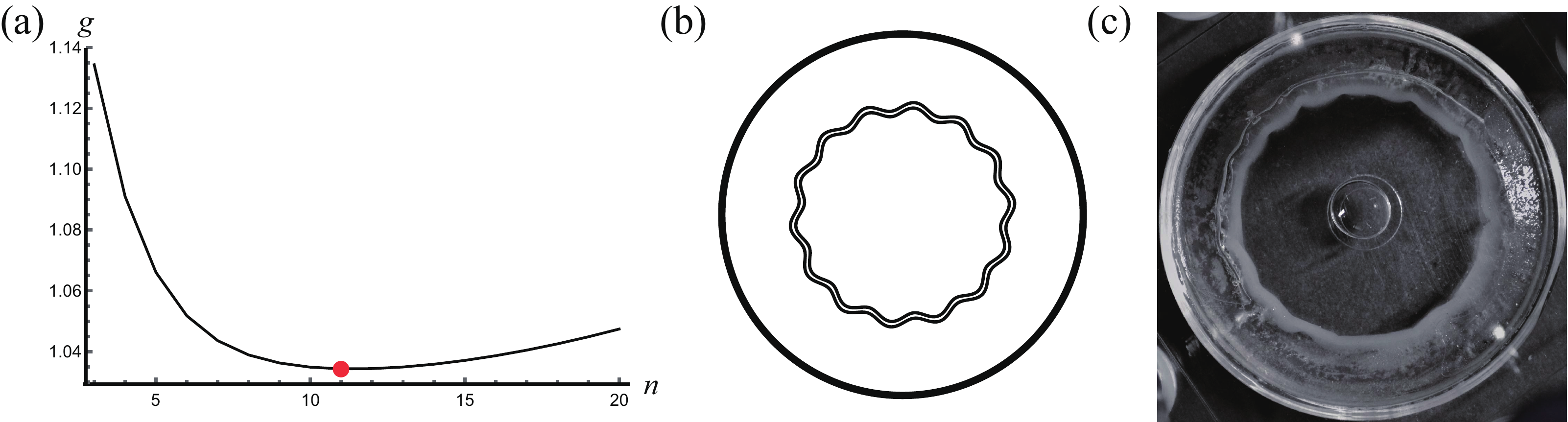}
    \caption{(Color online) Comparison of the wrinkled pattern between theoretical and experimental results for $A^*=0.67$, $B^*=0.7$, and $\beta=4.21$. (a) Bifurcation curve determined from the bifurcation condition (\ref{eq37}). (b) Eigen-shape of the wrinkled pattern where the amplitude is manually prescribed. (c) Wrinkled pattern in experiment.}
    \label{fig12}
\end{figure}

We assume that both the inner and outer layers are composed of incompressible neo-Hookean materials and the strain-energy function can be found in (\ref{eq1}). Similarly, the dimensionless parameters $A^*$, $B^*$, and $\beta=\mu/\hat{\mu}$ are used. Figure \ref{fig12} displays the bifurcation curves and the associated eigen-shape for $A^*=0.67$, $B^*=0.7$, and $\beta=4.21$, which are consistent with the material and geometrical parameters of sample $sp_1$. Meanwhile, a wrinkled pattern in experiment is shown for comparison. It is found that the bifurcation curve has a $U$-shape where the minimum identifies the first bifurcation point. The vertical coordinate of this minimum gives the critical growth factor $g_{cr}$ triggering surface wrinkling while the horizontal coordinate $n_{cr}$ counts the wavenumber. In detail, we obtain $g_{cr}=1.08738$ and $n_{cr}=14$. It is the growth factor $g$ that drives the deformation and further triggers surface an instability while the loading parameter is the chemical potential in swelling process. Furthermore, an accurate measurement of the thickness of the inner layer after deformation is difficult. For this reason, we no longer compare the critical load and only illustrate the corresponding comparisons for the wavenumber.

\begin{table}[h]
   \centering
    \caption{Comparisons of the wavenumber $n_{cr}$ between experiment and theoretical analyses.}
    \bigskip
    \begin{tabular}{cccccc}
   \hline label&$A^*$&$B^*$&modulus ratio &experiment&theory\\
     \hline$sp_1$&0.67&0.7&4.21&14&14\\
     \hline$sp_3$&0.67&0.7&6.35&11&14\\    
       \hline
    \end{tabular}
    \label{table3}
\end{table}

We illustrate the comparisons of $n_{cr}$ between experiment and theory in Table \ref{table3}. It can be seen that a relatively good agreement is found for samples $sp_1$ and $sp_3$. However, for the other sample $sp_4$ (note that interfacial wrinkles was seen in $sp_2$ so it is excluded in this comparison), there exists a large error. Then we intend to seek a possible source that is responsible for the difference. Note that the dimensionless thickness of the inner layer for this sample is given by $H^*=1-A^*/B^*\approx0.0143$, which is extremely thin. Referring to the scaling laws derived in Jin el al. \cite{33}, we obtain $n_{cr}\approx O(1/(H^*\beta^{-1/3}))$ if there is no growth in the outer layer. According to this relation, either the thickness of the inner layer or the modulus ratio between the two layers can alter the wavenumber. From the material characterization test in Section 2 and the comparisons in Table \ref{table3}, it is speculated that the modulus ratio is relatively accurate. Furthermore, it can be seen that $n_{cr}$ is highly sensitive to the variation of $H^*$, and any marginal geometrical mismatch of the tubular mold would generate an inaccurate $H^*$. Specifically, for $H^*\approx0.0143$, a higher relative error would appear and further produce the discrepancy between theoretical and experimental results. In spite of this fact, the desired accordance in Table \ref{table3} still signifies the validation of the growth model in reproducing the wrinkled pattern in growing tubular tissues.

\section{Post-buckling evolution}
The post-buckling evolutions in the swelling experiments have been described in detail in Section 2. In this section, we shall carry out a post-buckling analysis using finite element analysis (FEA) in commercial software Abaqus \cite{80} and then to make a comparison with experimental findings. In order to incorprate a volumetric growth, we establish a growth model by writing Abaqus UMAT subroutine codes following the user guideline. In all FE simulations, the built-in Module ``Static, General'' in Abaqus and the four-node linear plane-strain hybrid elements (CPE4H) are used. Note that the meshing procedure naturally develops a geometrical imperfection in the inner surface as circle is approximated by a polygon, so any ancillary geometrical or physical defect is unnecessary in all FE models. In this way, a nonlinear analysis for growth-induced deformation can be realized. It is worth mentioning that we shall only present a comparison of pattern evolution for samples $sp_1$, $sp_3$, and $sp_5$.
\begin{figure}[!h]
	\centering\includegraphics[scale=0.6]{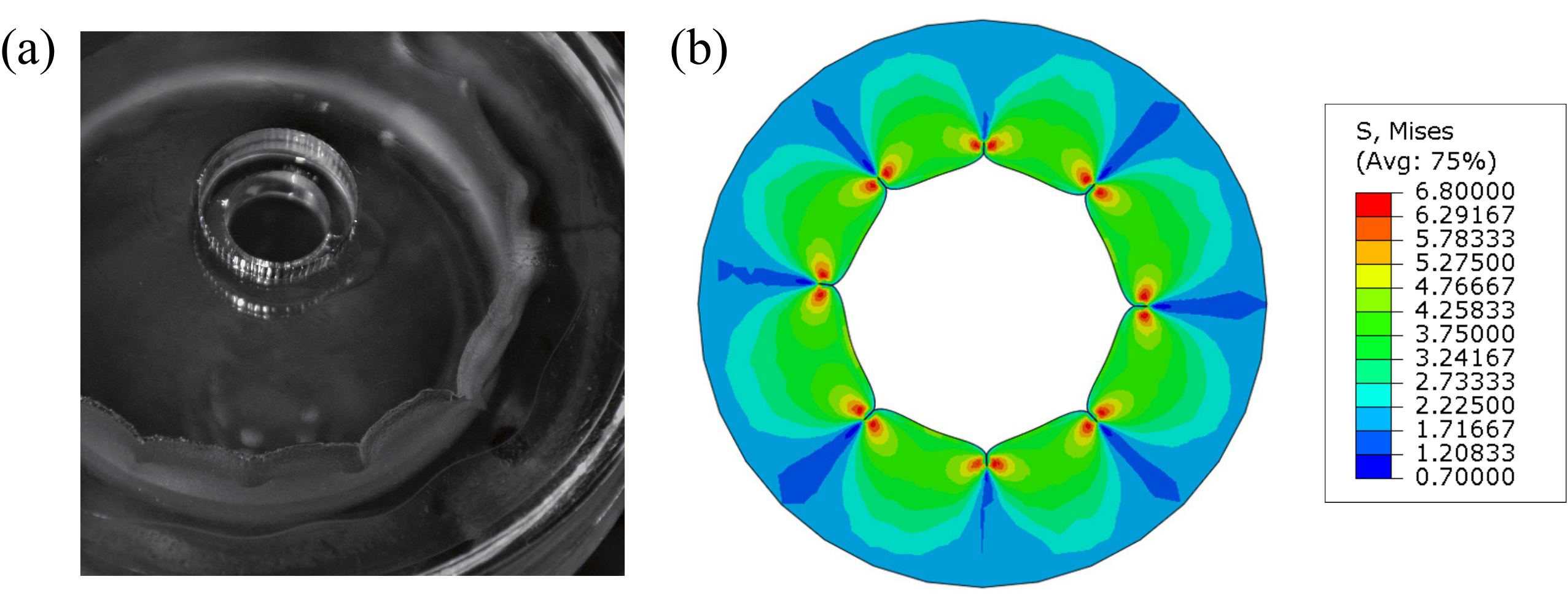}
    \caption{(Color online) Comparison of the creasing morphology for $A^*=0.67$, $B^*=0.7$, and $\beta\approx1$. (a) Picture taken from our swelling experiment. (b) Snapshot based on FE simulation accompanied by the legend of stress.}
	\label{fig13}
\end{figure}

Figure \ref{fig13} illustrates the surface patterns of experimental and FE results for the specimen $sp_5$ where the inner and outer layers share almost the same shear modulus. In the FE simulation, we take $\beta=1.1$. Since the shear moduli of the inner and outer layers are very close, there is also no surface wrinkling in our simulation and a highly localized creasing mode arises in the inner surface instead. In accordance with our experimental findings, a miniature crease turns into a deep crease as growth keeps going and no other pattern has been found. 

\begin{figure}[!h]
	\centering\includegraphics[scale=0.6]{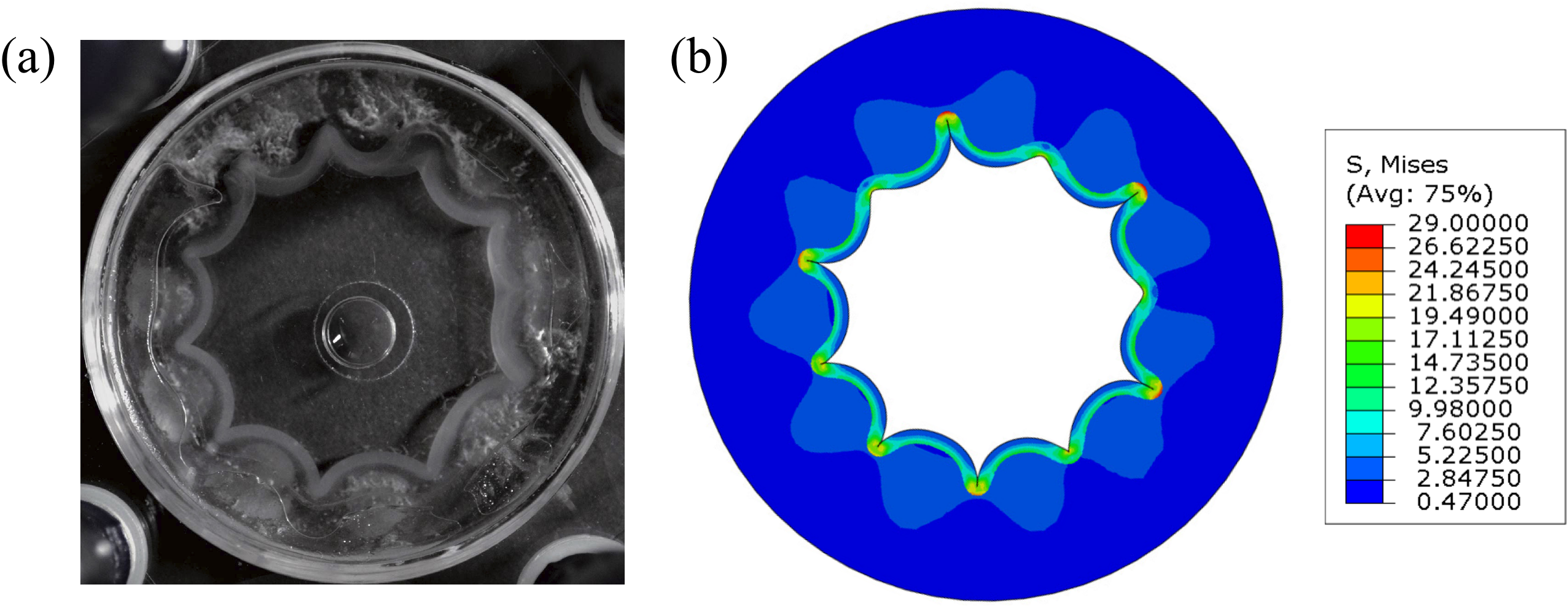}
    \caption{(Color online) Comparison of the creasing morphology between experimental observation and FE simulation. (a) The parameters are given by $A^*=0.67$, $B^*=0.7$, and $\beta\approx6.35$. (b) The parameters are given by $A^*=0.662$, $B^*=0.7$, and $\beta=6.35$.}
	\label{fig14}
\end{figure}

Figure \ref{fig14} plots the creasing morphology for sample $sp_3$ following progressive surface wrinkles for both experimental and FE outcomes. In this case, every wrinkle will deepen until a self-contact occurs and then form a creasing pattern. So the number of crease is consistent with the number of wrinkles. It can be seen from Table \ref{table3} that the wavenumber based on theoretical prediction is inconsistent with the counterpart counted in the experiment. Below Table \ref{table3} we have demonstrated a possible source giving rise to the inaccuracy, which is the fabrication mismatch of the geometrical size, especially the thickness of the inner layer. To provide qualitative insight into the evolution of surface wrinkling, we slightly amend the inner radius of the structure by $A^*=0.662$ in our FE model while other parameters remain the same, and this value is selected since it corresponds to $n_{cr}=11$ based on our theoretical model. Recall that the original $A^*$ is 0.67, thus the relative error between these two values is around $1.2\%$. This again confirms that a minor vibration of the inner layer thickness can induce a relatively great inaccuracy of the wavenumber. Seen from Figure \ref{fig14}, the two morphology are quite similar, which offers a further validation of the growth model in capturing essential features of pattern evolution. 

\begin{figure}[!h]
	\centering\includegraphics[scale=0.43]{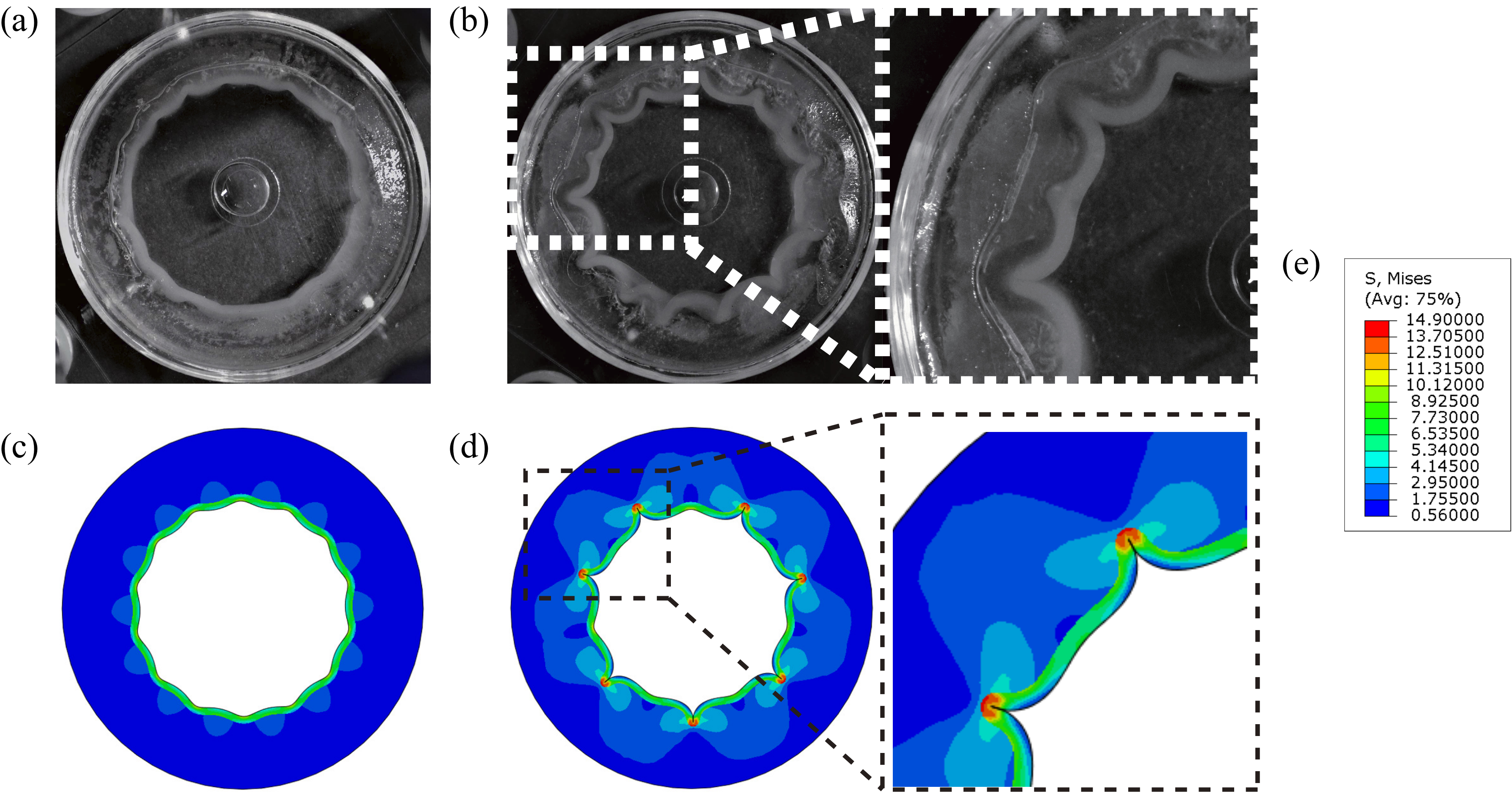}
	\caption{(Color online) Comparison of the creasing morphology between experimental observation and FE simulation for $A^*=0.67$, $B^*=0.7$, and $\beta=4.21$. (a) A wavy pattern from swelling experiment. (b) Period-doubling profile from swelling experiment. (c) Surface wrinkling based on FE simulation. (d) Period-doubling mode according to FE simulation. (e) The accompanied legend of stress.}
	\label{fig15}
\end{figure}

Finally, we exhibit the comparisons of pattern evolution of $sp_1$ in Figure \ref{fig15}, where all parameters of the FE model are identical to those used in our experiment. In this condition, our FE calculation yields the critical growth factor and the associated wavenumber as $g_{cr}=1.08735$ and $n_{cr}=14$, respectively, which agree extremely well with the counterparts based on theoretical model. It turns out that a sinusoidal shape with 14 wrinkles is observed in our experimental, theoretical, and FE results. Furthermore, as growth continues, the amplitude of wrinkle increases. At another critical value of growth, a period-doubling phenomenon is set off where each period involves a wrinkle and a crease. This special profile is recorded by our experiment as well as FE simulation, see Figure \ref{fig15}. In particular, because the wavenumber is an even integer, a perfect period-doubling can be observed in FE simulations. However, in our swelling experiment, a variation of either the material property or the sample size may affect pattern evolution. This explains why some places have formed period-doubling mode but some positions have not. Nevertheless, we still believe that the FE model captured the critical information of pattern transition in growing tubular tissues.

From these comparisons, we conclude that the FE simulations based on the volumetric growth in \cite{18} can qualitatively describe all deformation features, including the initiation of a specific mode, evolution of the surface profile, and transition between different patterns. 

\section{Conclusion}
The validity of a popular volumetric growth model in \cite{18} to imitate growth-induced pattern formation in growing bilayered tubular organs was verified by our experiments. We employed PDMS elastomer to fabricate simulacrums and designed a swelling experiment to mimic constrained growth of bilayered tubular tissues exploiting the fact that PDMS absorbs n-Hexanes. To experimentally unveil pattern formation induced by swelling/growth, we adopted the incompressible neo-Hookean model and characterized the material constant (shear modulus) using simple tension and simple shear. In doing so, three representative values of modulus were identified for PDMS samples with distinct mass ratios of base monomer to cross linker, and five samples were further prepared for swelling experiment. Especially, the calibrated data allow us to establish a twin model (the geometries and physical properties are the same as those for the samples) in theoretical and finite element (FE) models. From our swelling experiments, it is found that creasing or wrinkling may occur as a result of primary bifurcation, depending on the value of the modulus ratio $\beta$. In addition, creasing instability never generates a mode transformation leading to other surface patterns. However, not only the transition from wrinkle to period-doubling mode but also the transition from wrinkle to creasing mode was observed in our experiments. In particular, the period-doubling mode is a combination of wrinkles and creases. We emphasize that the experimental findings with regard to the effect of geometrical and material parameters on the onset of surface wrinkling and the associated wavenumber were qualitatively consistent with the conclusions in previous studies \cite{24,25,26,33}. Also, our experimental findings furnish a further evidence to the viewpoint that the modulus ratio $\beta$ acts a chief role in governing mode transition \cite{62,63}.

Then we established a theoretical model based on the growth theory by Ben Amar and Goriely \cite{18}. Specifically, referring to our earlier work in \cite{43}, a homogeneous growth model was employed. Compared with the studies by Li et al. \cite{24} and Jin et al. \cite{33}, a buckling analysis using Stroh formulation and impedance matrix method was carried out. We have amended the surface impedance matrix method such that it still works for displacement boundary conditions. It turns out that the bifurcation analysis based on Stroh method is more compact, especially for multilayer structures. Finally, a FE model was built in Abaqus by virtue of UMAT subroutine coding, and a post-buckling analysis was performed to trace pattern evolution. In general, the theoretical predictions coincide with our experimental results for the buckling pattern and associated wavenumber, while the FE simulations are identical to pattern transitions that occurred in experiments. It is expected that our investigation would offer useful experimental insight into morphological formation in growing tubular tissues and further support the viewpoint that a phenomenological growth can capture the major features of surface instabilities and the corresponding pattern evolutions. Finally, we emphasize that an interfacial wrinkling instability was observed in our experiments. This distinctive instability scenario can not be explained by the theoretical model and the FE model in this paper. Some further investigations are necessary for this topic.

\section*{Acknowlegments}
This work was supported by the National Natural Science Foundation of China (Project Nos 12072227 and 12021002). The Abaqus simulations were carried out on TianHe-1 (A) at the National Supercomputer Center in Tianjin, China. We thank Prof. Yibin Fu from Keele University for valuable discussions.


\begin{thebibliography}{100}

\bibitem{1}
Oh, JY, Kim, JY, Chan, WP, et al.
\newblock Spontaneously formed wrinkled substrates for stretchable electronics using intrinsically rigid materials.
\newblock {\em IEEE Electr. Device L.} 2016; 37(5): 588--590.

\bibitem{2}
Huang, YA, Ding, YJ, Bian, J, et al.
\newblock Hyper-stretchable self-powered sensors based on
  electrohydrodynamically printed, self-similar piezoelectric nano/microfibers.
\newblock {\em Nano Energy} 2017; 40: 432--439.

\bibitem{3}
Wu, HS, Kustra, S, Gates, EM, et al.
\newblock Topographic substrates as strain relief features in stretchable organic thin film transistors.
\newblock {\em Org. Electron.} 2013; 14(6): 1636--1642.

\bibitem{4}
Zhang, Y, Huang, Y, and Rogers, J.A.
\newblock Mechanics of stretchable batteries and supercapacitors.
\newblock {\em Curr. Opin. Solid St. M.} 2015; 19(3): 190--199.

\bibitem{5}
Bayat, A, and Gordaninejad, F.
\newblock Switching band-gaps of a phononic crystal slab by surface instability.
\newblock {\em Smart Mater. Struct.} 2015; 24(7): 075009.

\bibitem{6}
Sabbah, A, Youssef, A, and Damman, P.
\newblock Superhydrophobic surfaces created by elastic instability of PDMS.
\newblock {\em Appl. Sci.} 2016; 6(5): 152.

\bibitem{7} Dai, H-H, and Liu, Y. Critical thickness ratio for buckled and wrinkled fruits and vegetables. {\em Europhys. Lett.} 2014; 108: 44003.

\bibitem{8} Tallinen, T, Chung JY, Biggins JS, et al. Gyrification from constrained cortical expansion. {\em P. Natl. Acad. Sci. USA} 2014; 111(35): 12667--12672 .

\bibitem{9} Ambrosi, D, Ben Amar, M, Cyron, CJ, et al. Growth and remodelling of living tissues: perspectives, challenges and opportunities. {\em J. R. Soc. Interface} 2019; 16: 20190233.

\bibitem{10} Balbi, V, Destrade, and M, Goriely, A. Mechanics of human brain organoids. {\em Phys. Rev. E} 2020; 101: 022403.

\bibitem{11}
Shyer, AE, Tallinen, T, Nerurkar, NL, et al.
\newblock Villification: How the gut gets its villi.
\newblock {\em Science} 2013; 342(6155): 212--218.

\bibitem{12}
Budday, S, Raybaud, C, and Kuhl, E.
\newblock A mechanical model predicts morphological abnormalities in the developing human brain.
\newblock {\em Sci. Rep.} 2014; 4: 5644.

\bibitem{13}
Ben Amar, M, Chatelain, C, and Ciarletta, P.
\newblock Contour instabilities in early tumor growth models.
\newblock {\em Phys. Rev. Lett.} 2011; 106(14): 148101.

\bibitem{14}
Eskandari, M, Pfaller, MR, and Kuhl, E.
\newblock On the role of mechanics in chronic lung disease.
\newblock {\em Materials} 2013; 6(12): 5639--5658.

\bibitem{15}
Budday, S, Steinmann, P, and Kuhl, E.
\newblock The role of mechanics during brain development.
\newblock {\em J. Mech. Phys. Solids} 2014; 72: 75--92.

\bibitem{16}
Goriely, A, Vandiver, R, and Destrade, M.
\newblock Nonlinear Euler buckling.
\newblock {\em Proc. R. Soc. A} 2008; 464: 3003--3019.

\bibitem{17}
Rodriguez, EK, Hoger, A, and Mcculloch, AD.
\newblock Stress-dependent finite growth in soft elastic tissues.
\newblock {\em J. Biomech.} 1994; 27(4): 455--467.

\bibitem{18}
Ben Amar, M, and Goriely, A.
\newblock Growth and instability in elastic tissues.
\newblock {\em J. Mech. Phys. Solids} 2005; 53(10): 2284--2319.

\bibitem{19} Li B, Cao Y, Feng X, et al. 2012 Mechanics of morphological instabilities and surface wrinkling in soft materials: a review, Soft Matter 8, 5728--5745.

\bibitem{20} Goriely A. 2017 The Mathematics and Mechanics of Biological Growth, Springer-Verlag New York.

\bibitem{21}
Seow, CY.
\newblock Response of arterial smooth muscle to length perturbation.
\newblock {\em J. Appl. Physiol.} 2000; 89: 2065--2072.

\bibitem{22}
Balbi, V, Kuhl, E, and Ciarletta, P.
\newblock Morphoelastic control of gastro-intestinal organogenesis: Theoretical predictions and numerical insights.
\newblock {\em J. Mech. Phys. Solids} 2015; 78: 493--510.

\bibitem{23}
Eskandari, M, Javili, A, and Kuhl, E.
\newblock Elastosis during airway wall remodeling explains multiple co-existing instability patterns.
\newblock {\em J. Theor. Biol.} 2016; 403: 209--218.

\bibitem{24}
Li, B, Cao, YP, Feng, XQ, et al.
\newblock Surface wrinkling of mucosa induced by volumetric growth: Theory, simulation and experiment.
\newblock {\em J. Mech. Phys. Solids} 2011; 59(4): 758--774.

\bibitem{25}
Li, B, Cao, YP, Feng, XQ. Growth and surface folding of esophageal mucosa: A biomechanical model. {\em J. Biomech.} 2011; 44: 182--188.

\bibitem{26}
Moulton, DE, and Goriely, A.
\newblock Possible role of differential growth in airway wall remodeling in asthma.
\newblock {\em J. Appl. Physiol.} 2011; 110(4): 1003--1012.

\bibitem{27}
Moulton, DE, and Goriely, A.
\newblock Circumferential buckling instability of a growing cylindrical tube.
\newblock {\em J. Mech. Phys. Solids} 2011; 59(3): 525--537.

\bibitem{28}
Ciarletta P, and Ben Amar M. Growth instabilities and folding in tubular organs: a variational method in non-linear elasticity. {\em Int. J. Non-Linear Mech.} 2012; 47: 248--257.

\bibitem{29}
Ciarletta P, and Ben Amar M. Pattern formation in fiber-reinforced tubular tissues: folding and segmentation during epithelial growth. {\em J. Mech. Phys. Solids} 2012; 60: 525--537. 

\bibitem{30}
Balbi, V, and Ciarletta, P.
\newblock Morpho-elasticity of intestinal villi.
\newblock {\em  J. R. Soc. Interface} 2013; 10(82): 20130109.

\bibitem{31}
Ciarletta, P, Balbi, V, and Kuhl, E.
\newblock Pattern selection in growing tubular tissues.
\newblock {\em Phys. Rev. Lett.} 2014; 113(24): 248101.

\bibitem{32}
Balbi, V, Kuhl, E, and Ciarletta, P. Morphoelastic control of gastro-intestinal organo-genesis: Theoretical predictions and numerical insights. {\em J. Mech. Phys. Solids.} 2015; 78: 493--510.

\bibitem{33}
Jin, L, Liu, Y, and Cai, Z.
\newblock Asymptotic solutions on the circumferential wrinkling of growing tubular tissues.
\newblock {\em Int. J. Eng. Sci.} 2018; 128: 31--43.

\bibitem{34}
Jin, L, Liu, Y, and Cai, Z.
\newblock Post-buckling analysis on growing tubular tissues: A semi-analytical approach and imperfection sensitivity.
\newblock {\em Int. J. Solids Struct.} 2019; 162: 121--134.

\bibitem{35}
Liu Y, Zhang, Z, Devillanova, G, et al. Surface instabilities in graded tubular tissues induced by volumetric growth, {\em Int. J. Non-linear Mech.} 2020; 127: 103612.

\bibitem{36}
Sultan, E, and Boudaoud, A. The buckling of a swollen thin gel layer bound to a compliant substrate. {\em J. Appl. Mech.} 2008; 75: 051002.

\bibitem{37}
Ben Amar, M, and Ciarletta, P. Swelling instability of surface-attached gels as a model for soft tissue growth under geometric constraints. {\em J. Mech. Phys. Solids} 2010; 58: 935--954.

\bibitem{38}
Tokarev, I, and Minko, S.
\newblock Stimuli-responsive hydrogel thin films.
\newblock {\em Soft Matter} 2009; 5(3): 511--524.

\bibitem{39}
Dervaux, J, Couder, Y, Ben Amar, M, et al.
\newblock Shape transition in artificial tumors: from smooth buckles to singular creases.
\newblock {\em Phys. Rev. Lett.} 2011; 107(1): 018103.

\bibitem{40}
Tallinen, T, Chung, JY, Rousseau F, et al. On the growth and form of cortical convolutions. {\em Nat. Phys.} 2016; 12: 588--593.

\bibitem{41}
Holland, M, Budday, S, Goriely, A, et al.
\newblock Symmetry breaking in wrinkling patterns: Gyri are universally thicker than sulci.
\newblock {\em Phys. Rev. Lett.} 2018; 121: 228002.

\bibitem{42}
Du, Y, L\"u, C, Liu, C, et al.
\newblock Prescribing patterns in growing tubular soft matter by initial residual stress.
\newblock {\em Soft Matter} 2019; 15: 8468--8474.

\bibitem{43}
Liu, RC, Liu, Y, and Cai, Z.
\newblock Influence of the growth gradient on surface wrinkling and pattern transition in growing tubular tissues.
\newblock {\em Proc. R. Soc. A} 2021; 477: 20210441.

\bibitem{44}
Fu, YB, Liu, JL, and Franciso, GS. Localized bulging in an inflated cylindrical tube of arbitrary thickness - the effect of bending stiffness. {\em J. Mech. Phys. Solids.} 2016; 90: 45--60.

\bibitem{45}
Liu Y. Axial and circumferential buckling of a hyperelastic tube under restricted compression. {\em Int. J. Non-linear. Mech.} 2018; 98: 145--153.

\bibitem{46}
Stroh AN. Steady state problems in anisotropic elasticity. {\em J. Math. Phys.} 1962; 41: 77--103.

\bibitem{47}
Biryukov, SV. Impedance method in the theory of elastic surface waves. {\em Sov. Phys. Acoust.} 1985; 31: 350--354.

\bibitem{48}
Fu, Y, and Mielke, A. A new identity for the surface impedance matrix and its application to the determination of surface-wave speeds. {\em Proc. R. Soc. A} 2002; 458: 2523--2543.

\bibitem{49}
Fu YB. An integral representation of the surface-impedance tensor for incompressible elastic materials. {\em J. Elasticity} 2005; 81(1): 75--90.

\bibitem{50}
Ciarletta, P, and Destrade, M.
\newblock Torsion instability of soft solid cylinders.
\newblock {\em IMA J. Appl. Math.} 2014; 79(5): 804--819.

\bibitem{51}
Su, Y, Zhou, W, Chen, W, et al.
\newblock On buckling of a soft incompressible electroactive hollow cylinder.
\newblock {\em Int. J. Solids Struct.} 2016; 97-98: 400--416.

\bibitem{52}
Su, Y, Wu, B, Chen, W, et al.
\newblock Finite bending and pattern evolution of the associated instability for a dielectric elastomer slab.
\newblock {\em Int. J. Solids Struct.} 2019; 158: 191--209.

\bibitem{53}
Su, Y.
\newblock Voltage-controlled instability transitions and competitions in a finitely deformed dielectric elastomer tube.
\newblock {\em Int. J. Eng. Sci.} 2020; 157: 103380.

\bibitem{54}
Lee, JN, Park, C, and Whitesides, GM.
\newblock Solvent compatibility of poly(dimethylsiloxane)-based microfluidic devices.
\newblock {\em Anal. Chem.} 2003; 75(23): 6544--6554.

\bibitem{55}
Liu, M, and Chen, Q.
\newblock Characterization study of bonded and unbonded polydimethylsiloxane aimed for bio-micro-electromechanical systems-related applications.
\newblock {\em J. Microlith. Microfab.} 2007; 6(2): 023008.

\bibitem{56}
Kim, M, Moon, BU, and Hidrovo, CH.
\newblock Enhancement of the thermo-mechanical properties of PDMS molds for the hot embossing of PMMA microfluidic devices.
\newblock {\em J. Micromech. Microeng.} 2013; 23(9): 095024.

\bibitem{57}
Cai, Z, and Fu, Y. On the imperfection sensitivity of a coated elastic half-space. {\em Proc. R. Soc. Lond. A.} 1999; 455: 3285--3309.

\bibitem{58}
Hutchinson, JW. The role of nonlinear substrate elasticity in the wrinkling of thin films. {\em Phil. Trans. R. Soc. A.} 2013; 371: 20120422.

\bibitem{59}
Chen, X, and Hutchinson, JW.
\newblock {Herringbone Buckling Patterns of Compressed Thin Films on Compliant Substrates}.
\newblock {\em J. Appl. Mech.} 2004; 71(5): 597--603.

\bibitem{60}
Liu, Y, and Dai, H-H. Compression of a hyperelastic layer-substrate structure: Transitions between buckling and surface modes. {\em Int. J. Eng. Sci.} 2014; 80: 74--89.

\bibitem{61}
Fu, YB, and Cai, ZX. An asymptotic analysis of the period-doubling secondary bifurcation in a film/substrate bilayer. {\em SIAM J. Appl. Math.} 2015; 75: 2381--2395.

\bibitem{62}
Wang, Q, and Zhao, X. A three-dimensional phase diagramof growth-induced surface instabilities. {\em Sci. Rep.} 2015; 5: 8887.

\bibitem{63}
Zhao, R, and Zhao, X. Multimodal surface instabilities in curved film-substrate Structures. {\em J. Appl. Mech. ASME} 2017; 84: 081001.

\bibitem{64}
Cai, ZX, and Fu, YB. Effects of pre-stretch compressibility and material constitution on the period-doubling secondary bifurcation of a film/substrate bilayer. {\em Int. J. Non-Linear Mech.} 2019; 115: 11--19.

\bibitem{65}
Johnston, ID, McCluskey, DK, Tan, CKL, et al.
\newblock Mechanical characterization of bulk sylgard 184 for microfluidics and microengineering.
\newblock {\em J. Micromech. Microeng.} 2014; 24(3): 035017.



\bibitem{66}
Jiang, M, Lawson, ZT, Erel, V, et al.
\newblock Clamping soft biologic tissues for uni-axial tensile testing: A brief survey of current methods and development of a novel clamping mechanism.
\newblock {\em J. Mech. Behav. Biomed.} 2020; 103: 103503.

\bibitem{67}
Pucci, E, and Saccomandi, G. A note on the gent model for rubber-like materials. {\em Rubber Chem. Technol.} 2002; 75: 839--852.

\bibitem{68}
Ogden, RW, Saccomandi, G, and Sgura, I. Fitting hyperelastic models to experimental data. {\em Comput. Mech.} 2004; 34: 484--502.

\bibitem{69}
Mihai, LA, and Goriely A. How to characterize a nonlinear elastic material? A review on nonlinear constitutive parameters in isotropic finite elasticity. {\em Proc. R. Soc. A} 2017; 473: 20170607.

\bibitem{70}
Ogden RW.
\newblock Non-linear elastic deformations.
\newblock {\em Dover Civil and Mechanical Engineering} 1997.

\bibitem{71}
Roucou, D, Diani, J, Brieu, M, et al.
\newblock Critical strain energy release rate for rubbers: single edge notch tension versus pure shear tests.
\newblock {\em Int. J. Fracture} 2019; 216: 31¨C-39.

\bibitem{72}
Wang Z, Volinsky, AA, and Gallant, ND. Crosslinking effect on polydimethylsiloxane elastic modulus measured by custom-built compression instrument. {\em J. Appl. Polym. Sci.}  2014; 131: 41050.

\bibitem{73}
Chen, Z, Zhang, X, and Song, J. Surface wrinkling of an elastic graded layer. {\em Soft Matter}  2018; 14: 8717.

\bibitem{74}
Jin, L, Chen, D, Hayward, RC, et a. Creases on the interface between two soft materials. {\em Soft Matter} 2014; 10: 303--311.

\bibitem{75}
Razavi, MJ, Pidaparti, R. and Wang, X. Surface and interfacial creases in a bilayer tubular soft tissue. {\em Phys. Rev. E} 2016; 94: 022405.

\bibitem{76}
Fu, YB, and Ogden, RW.
\newblock Nonlinear stability analysis of pre-stressed elastic bodies.
\newblock {\em Continuum Mechanics and Thermodynamics} 1999, 11(3): 141--172.

\bibitem{77}
Shuvalov, AL. A sextic formalism for three-dimensional elastodynamics of cylindrically anisotropic radially inhomogeneous materials. {\em Proc. R. Soc. Lond. A} 2003; 459: 1611--1639.

\bibitem{78}
Shuvalov, AL. The Frobenius power series solution for cylindrically anisotropic radially inhomogeneous elastic materials. {\em Q. J. Mech. Appl. Math.} 2003; 56: 327--345.

\bibitem{79}
Wolfram Research Inc. Mathematica: version 12. Wolfram Research Inc. Champaign, IL. 2019.

\bibitem{80}
\newblock  ABAQUS Analysis User¡¯s Manual, version 6.13, Dassault Syst¨¨mes. Providence, RI, USA 2013.





\end{thebibliography}
 \end{document}